# Mysterious Role of Cap Configuration in Single-Walled Carbon Nanotube Catalytic Growth

Tianliang Feng, Ziwei Xu*

*School of Materials Science and Engineering, Jiangsu University, Zhenjiang 212013, China*

## Abstract

Understanding the role of cap structure during the nucleation and growth of single-walled carbon nanotubes (SWCNTs) is essential for achieving chirality-controlled synthesis. In this work, we propose a novel and intuitive algorithm to determine the chirality of nascent carbon caps by tracking the relative shifts of six pentagons within a topological coordinate system. Based on this algorithm, we propose three routes of pentagon shifts for the chirality mutations from armchair (AC) to near-AC caps, namely the transverse shift (n,n) → (n+1,n-1), inward shift (n,n) → (n-1), and outward shift (n,n) → (n+1,n), can occur according to the energy profiles calculated based on the density function theory (DFT), providing a new perspective to explain the experimental abundance of near-AC SWCNTs. After that, we construct 24 representative short caps and long SWCNTs with different chiralities and symmetries, and perform DFT calculations to evaluate their thermodynamic stability and deformation behaviors on flat Ni(111) surfaces and curved $Ni_{55}$ particle, respectively. The results reveal that cap topology and catalyst curvature together induce dual deformations that significantly influence the formation and interface energies. Notably, AC and near-AC caps exhibit superior flexibility and robust energetic advantages, especially the (6,6) cap with $C_{6v}$ symmetry. A linear correlation between cap-induced deformation and formation energy is established through a defined shape factor, highlighting the interaction between cap structure and catalyst interface as a key driver of chirality selection. This study theoretically reveals the cap evolution mechanism and lays the foundation for the rational design of SWCNT growth strategies.

* Corresponding Author. E-mail address: ziweixu2014@ujs.edu.cn (Z. Xu)

## 1. Introduction

Since their discovery in 1993,[1,2] single-walled carbon nanotubes (SWCNTs) have garnered considerable attention due to their multiple advantages as unique one-dimensional topological material, including exceptionally ultra-high mechanical strength, excellent electronic conductivity, and remarkable thermal stability.[3-6] These characteristics endow SWCNTs with immense potential across numerous fields, encompassing photoelectric detection,[7-9] biological and chemical sensing,[10-17] and conductive films,[18-20] thereby holding profound significance for both scientific exploration and industrial production.[21,22] However, some applications remain elusive due to the challenges in achieving precise control during SWCNT synthesis.

SWCNTs are obtained as mixtures of numerous chiral species with various electronic properties, which complicates their potential application in the field of electronics.[23,24] For large-scale, high-performance SWCNT-based electronic applications, such as high-performance field-effect transistors (FETs),[25-28] the ultra-high semiconductor-type SWCNT purity required (>99.9999%)[29,30] is crucial for digital electronics and integrated circuits (ICs).[31] Such purity can only be achieved through complex post-sorting processes and cannot yet be directly attained via catalytic growth. Phenomena such as bundling, stacking, and disordered aggregation that may occur will all increase the complexity of assembly and device integration processes.[32] This limits the fabrication of high-performance SWCNT FETs, as even trace amounts of metallic SWCNTs within the transistor channel may result in high leakage currents and erroneous logic functionality.[33,34] In addition to purity, high-precise chirality control is crucial for consistent device performance, including threshold voltage, temperature sensitivity, and contact resistance.[35] During the sustained attempts on the selective synthesis of SWCNTs, many experiments found the preferred abundance of armchair (AC) or near-AC SWCNTs during the selective growth of SWCNTs.[36-45] In 2022, Nishihara and Miyauchi et al. found that the abundance distribution of SWCNT with a chiral angle ($\theta$) of about 20° increased anomalously with increasing angle, statistically confirming the preferential growth of AC SWCNTs.[46]

Carbon cap nucleation is the primary key step determining the initial chirality in the catalytic synthesis of SWCNTs. In the nucleation stage, once upon the arrangement of the six pentagons is fixed, the nascent carbon cap has already encoded the chirality of the subsequent SWCNT's wall (assuming that no more defect introduced) despite the fact that the wall has not been formed. Just like the "chirality gene" in SWCNT, we cannot "see" the chirality intuitively from the cap edge or the helicity of the SWCNT wall during this nascent cap stage. Since 1999, numerical algorithms for enumerating the caps for specific chirality have been developed by Brinkmann et al.[47,48] and Astakhova et al.[49,50] independently based on the projection on the honeycomb lattice used in fullerene.[51] It was found that the number of caps ($N^{cap}$) increase rapidly with diameters of SWCNTs ($d^{\gamma}$), namely, $N^{cap} \sim d^{\gamma}$, where $\gamma \approx 8$.[47] Reich et al. conducted a detailed study of carbon nanocaps, mainly in terms of their geometry, structure, and energetics, and combined chirality with formation energies to propose a method for constructing nanotube caps by cutting 60 ° cones from a hexagonal lattice.[52] This study suggested that the carbon cap structure uniquely determines the choice of SWCNT chirality and a pair of adjacent pentagons requires a much larger formation energy ($E_f$) of ~ 1.5 eV. Murr et al. outlined an easy-to-apply graphical modeling technique for the rapid creation of any type of AC, Zigzag (ZZ), chiral or achiral defect-free nanotubes.[53]

Robinson et al. proposed a general method to generate the caps based on the approach of energy minimization in *The Thomson Problem*,[54] which can guarantee the caps with higher physical reasonability and the lowest energy. In 2017, Sugai et al. developed a numerical algorithm and package to realize the two-way correspondence between carbon nanotubes and caps. And, the caps designed contain various polygons from triangles to heptagons.[55] However, most of these methods originate from the lattice projection or graph theory. For a given nascent cap that already has a fixed arrangement of six pentagons, but the helicity of the tube wall is not yet clear, it is rather difficult to see the corresponding chirality directly, unless one resorts to the map of the lattice projection that lacks intuitiveness. Also, some of these developed algorithms are limited by geometrical precondition that leads to some omitted structures. For instance, due to the inhomogeneous projection, those caps far from the sphere shape could not be found based on the *Thomson* method[54]. In addition, in our previous study, given a nascent cap with fixed pentagon distribution, we proposed that just by the movement of the 6$^{th}$ pentagon along the edge of the carbon cap, it is possible to traverse almost all the chiralities corresponding to the carbon cap diameter from AC to ZZ on the liquid catalyst. Nevertheless, if more pentagon's movement has been considered, the caps variety for a certain chirality will be greatly increased.

In addition to these developed numerical methods, the possible effect of the cap structure on the catalytic growth of SWCNT has been extensively studied. As early as 2006, Reich et al. proposed epitaxial growth of SWCNT dominated by the geometrical lattice matching between the cap and catalyst surface.[56] Chen et al. suggested that precise management of the concentration of individual C atoms and C2 dimers in the experimental environment at different reaction stages could potentially enable chiral selective growth of SWCNTs.[57] Subsequently, Balbuena et al. observed a slight increase in the tendency to form near-AC shaped SWCNTs with the use of C(2) precursors.[58] Penev et al. performed a large-scale calculation to present an energy landscape of all possible isomers of carbon caps for chirality from AC to ZZ SWCNTs and conclude that there is no clear correlation between cap structure and $\theta$.[59] Later, Luo et al. further calculated the cap energies on different shapes of the metal catalyst and concluded the negligible effect of the catalyst morphology on the chirality selection.[60] Instead, they believe that the SWCNT-catalyst interface should be the dominant factor for the chirality bias.[59,60] Indeed, many theorical works have explained the abundance of AC or near-AC SWCNTs based on the thermodynamic stability, kinetic growth rate as well as the configuration entropy contribution of the SWCNT-catalyst interface.[61-65] Differently, Xu et al. proposed the kinetic assignment of the SWCNT chirality caused by the random formation of 6$^{th}$ (or say the last) pentagon. The control of the 6$^{th}$ pentagon's position on some specific catalysts with energetically favored chemical site is therefore proposed to control the "chirality gene" of the nucleated cap and subsequent SWCNT.[66] Recently, by considering the energy competition between the cap curvature and the interface energy, the growth transformation from (6,5) dominant SWCNTs at 700 °C to (7,5) dominant SWCNTs at 800 °C, which is observed in the experiments, has been well explained.[41]

Despite of these studies, a simple algorithm for enumerating the caps of all isomers and identify their chirality based on a given cap with determined pentagon arrangement is still lacking, especially for the one that can describe the chirality evolution by movements of any number of the six pentagons in the ready cap. Recently, the chirality mutation via the introduction of pentagon-heptagon (5|7) pair defect in the SWCNT wall has been proposed to explain the experimental abundance of (n,n-1) SWCNTs.[67,68] Zhu and Wei et al. recently proposed the concepts of symmetry and asymmetry evolution in SWCNT growth, attributed to the intragenerational and transgenerational mutations, respectively, which are atomically determined by the 5|7 defect conformation embedded within the carbon wall.[69] However, the feasibility of the similar chirality mutation via the evolution of the cap structure remains elusive during the nucleation stage of SWCNT. Moreover, rare attentions have been paid to the co-deformation induced by the cap structure and the interfaces and the associated influence on the SWCNT catalytic growth. Herein, in this theoretical study, we firstly proposed a new simple algorithm based on an analogy coordinate system, which enables direct enumeration and identification operations on a three dimensional cap. Based on this algorithm, 24 representative caps from AC to ZZ (each chirality has four cap isomers) have been built to investigate the dual deformation effect of these caps and their interfaces on the catalytic growth of SWCNT on the flat Ni(111) surface and curved $Ni_{55}$ particle. Through density functional theory calculations, we compared the energy landscapes and deformations of these caps and their elongated SWCNTs on the two catalyst surfaces. The results show that cap structures and the surface curvature can result in energetic differences. And, the dual-deformation caused by both the cap and the interface display high dependence on the chirality. The AC caps as well as their SWCNTs present robust energetic advantages and flexible deformation preferences during both the nucleation stage and the elongation stage, respectively. Finally, the chirality mutations from AC to near-AC were confirmed to be energetically feasible through the pentagon's movement during the nucleation stage, which may provide another view on the abundance of the AC or near-AC SWCNTs synthesized in the experiments. This study can further deepen our insights into the effect of the cap structure on the SWCNT catalytic growth and provide a theoretical guidance for the controlled synthesis of SWCNTs.

## 2. Materials and Methods

### 2.1 Density functional theory (DFT) calculations

In this work, all the first-principles calculations were carried out with spin-polarization considered based on the density functional theory (DFT) method as implemented in Vienna ab initio simulation package (VASP).[70] The exchange-correlation energy was described by using local density approximation (LDA)[71,72] with the projector-augmented wave (PAW) adopted.[73,74] The dispersion correction based on the DFT-D3 scheme was applied to describe the van der Waals (vdW) interactions. [75,76]

The capped SWCNT's models on the Ni(111) surface of two layers atoms and $Ni_{55}$

nanoparticle were adopted in the calculations with the cell sizes of 17.444 × 17.444 × 35 Å and 30 × 30 × 40 Å. A vacuum spacing of at least 15 Å between all structures was used to avoid the interactions of periodic images. A Γ-centered 2 × 2 × 1 k-point mesh was employed for all calculations. To reduce the large computational cost in this study, different parameter settings of energy cutoff of the planewave, energy convergence criteria of self-consistent calculation, force convergence criteria of ionic relaxation in different calculations, which depends on the system size and the required accuracy. For the calculations of pentagon shifts of the nascent caps and deformation energy on the Ni particle, the energy cutoff, energy convergence criteria, and force convergence criteria are set to be 400 eV, $10^{-4}$ eV, and 0.01 eV/Å, respectively. For the calculations of the short and long SWCNTs on the Ni(111) and $Ni_{55}$, however, they are set to be 250 eV, $10^{-3}$ eV, and 0.02 eV/Å, respectively. due to the substantial increases of computational cost and less required accuracy on the exact energies (i.e., focusing only the relative energies). Such a compromise strategy between the accuracy and computational efficiency has already been confirmed to be acceptable in previous DFT molecular dynamics simulations on the SWCNT growth on the alloy catalyst, which demands huge computational cost.[77]

Here, we considered six chiralities of SWCNTs for calculations, including (12,0), (11,1), (9,3), (8,4), (7,5), and (6,6). The $E_f$ for the SWCNT on the catalyst of Ni(111) or $Ni_{55}$ is defined by the equation (1):

$$E_f = \frac{E_{SWCNT/cat} - E_{cat} - N_C * \mu_C}{N_C} \qquad (1)$$

Where $E_{SWCNT/cat}$ represents the total energy of the SWCNT on the catalyst, $E_{\mathrm{cat}}$

represents the energy of catalyst, $N_c$ and $\mu_c$ denote the number of carbon atoms and the energy of one carbon atom in the graphene ($\mu_c$ = - 9.40 eV).

The interface formation energy ($E_{if}$) between the SWCNT edge and the attached is defined by the equation (2):

$$E_{if} = \frac{E_{SWCNT/cat} - E_{cat} - E_{SWCNT} - E_{ef}}{N_d} \quad (2)$$

Where $N_d$ is the number of the dangling carbon atoms at the edge of SWCNT and $E_{ef}$ is the edge formation energy of the SWCNT, which is defined by:

$$E_{ef} = \frac{E_{SWCNT} - 2E_{SWCNT/2}}{2N_d} \quad (3)$$

Where, $E_{SWCNT}$ is the energy of a long SWCNT and the $E_{SWCNT/2}$ is the energy of short SWCNT.

The interior energies ($E_{in}$) for each carbon atom is defined by:

$$E_{in} = \frac{E_f * N_c - E_{if} * N_d}{N_c - N_d} \quad (4)$$

## 2.2 Classic molecular dynamics simulations

The hybridized molecular dynamic and basin-hopping (H-MD-BH) simulation, developed in previous publication,[78] has been employed to simulate the nucleated caps on the nickel particles, such as $Ni_{19}$, $Ni_{32}$, and $Ni_{55}$. During the stage of MD simulation, the Velocity Verlet algorithm is applied to integrate the newton's equation of atom's motion with the timestep of 0.5 fs. The Berendsen thermostat is adopted to maintain the temperature ~ 1300 K.[79] During the stage of BH simulation, the Stone-Wales (SW) transformation[80] is stochastically performed with the probability of acceptance $p \sim e^{-\Delta E/k_B T}$, where $\Delta E$ is the energetic differences of the structure after and before the SW transformation. And, $k_B$ and $T$ are the Boltzmann constant and temperature

(T ~1300 K), respectively.

## 3. Results and discussion

### 3.1 New algorithm for identifying the cap chirality based on the basic coordinate system and pentagon shifts

We developed a new general algorithm to determine the chirality of the nascent cap based on the pentagon's relative positions, which can be applied to both the isolated-pentagon and non-isolated-pentagon caps. Firstly, like the mathematical coordinate systems in three-dimensional space, we define four basic coordinate systems (BCS) in the topological space of the carbon cap. As illustrated in Figure 1a, with a pentagon as the center (marked by five-pointed star), five extension lines (red dotted lines) are drawn perpendicular to the five edges of the center pentagon, and the other five pentagons that make up the carbon cap shifts along this extension line to form new carbon caps, all of which are the ZZ SWCNTs. This coordinate system is defined as the pentagon-Zigzag (Pen-ZZ) BCS. In this system, the smallest (5,0) cap is defined as the base-point cap (Figure 1b). The other chiralities derived from the shift of the other five pentagons along the five extension lines is $(\mathrm{n}_0, 0) + \sum_1^5 i * (1,0)$, where the $i$ is the number of the shift steps (a movement from one hexagon to adjacent hexagon in the graphene honeycomb lattice) of each pentagon along the extension line. As shown in Figure 1c, the chiralities of the newly formed ZZ caps derived from the base-point cap are (5,0) + (1,0) + (1,0) + (1,0) + (1,0) + (1,0) → (10,0), (5,0) + (1,0) +(1,0) + (1,0) + 2*(1,0) + 2*(1,0) → (12,0), and (5,0) + (1,0) +(1,0) + (1,0) +(1,0) + 3*(1,0) → (12,0), respectively. The Pen-AC BCS is defined by the five extension lines (blue solid lines) drawn from the five vertices of the center pentagon (Figure 1a). The base-point cap of the pentagon-Armchair (Pen-AC) BCS is (5,5) cap shown in Figure 1d. In contrast, all the caps with the five pentagons shifting on the extension lines of the Pen-AC BCS can grow AC SWCNTs. The chiralities can be expressed as $(5,5)+\sum_{1}^{5} i*(1,1)$. Here, the $i$ is the number of the shift steps from one hexagon to the next hexagon along the extension lines for each pentagon. Thus, the chiralities derived from the (5,5) base-point cap are shown in Figure 1e. Besides, the hexagon-Zigzag (Hex-ZZ) and hexagon-Armchair (Hex-AC) BCS as well as the chiralities of the caps on them can be defined by the same way (Figure 1f). The only difference is the base point cap is (6,0) cap and (6,6) cap for Hex-ZZ (Figure 1g) and Hex-AC (Figure 1i), respectively. Similarly, the chiralities derived from the (6,0) and (6,6) base-point caps are shown in Figure 1h,j, respectively.

Just like a coordinate system in mathematics, we can choose any of the four BCS as a reference to determine the final chirality of an arbitral cap through calculating the shift seps of the pentagons with respect to base-point cap. The chirality change at each shift step can be labeled according to the map in Figure 1k,l. In the frontal view of the target pentagon, there are two orientations of the shifting pentagon for the Pen (or Hex)-ZZ and Pen (or Hex)-AC systems, respectively, i.e., downward orientation (Figure 1k) and the upward orientation (Figure 1l), between which there is just a 30 ° rotation. Figure 1m shows a simple example of the chirality change with the shift of target pentagon in a (12,0) cap. The target pentagon is in a downward orientation with respect to the Pen-ZZ BCS. The chirality (12,0) will change by (-1,0), (-1,1), (0,1), (1,0), (1,- 1), and (0,-1) for one step of upward shift, upper right shift, lower right shift, downward shift, lower left shift, left shift, and upper left shift. As illustrated in Figure 1m, the continuous shift of the pentagon along one direction, say upper right direction, will lead to the continuous chirality change (-1,1), i.e., (12,0) → (11,1) → (10,2) → (9,3) → (8,4) → (7,5) → (6,6).

Figure S1 shows the more general examples to determine the chiralities of the nascent caps of (10,0), (9,2), (8,3), and (6,6) chiralities based on the pentagon shift. As shown in Figure S1a, we can select the Hex-ZZ BCS to determine the corresponded chirality of the target cap. Firstly, we draw the (6,0) base-point cap of Hex-ZZ BCS, where the center hexagon is marked by a red pentagram, surrounded by six pentagons (four blue-filled small pentagons and two orange-filled pentagons). Compared with the final six pentagons, two remain in their original positions (orange-filled), while the remaining four undergo shift (positions undergoing the transition from pentagon to hexagon, purple-filled), indicated by green arrows. The chirality change for each shift step can be labeled according to Figure 1g. The real chirality of this cap is therefore equal to the base-point chirality (6,0) + the total chirality changes created by all steps of the pentagons' shifts, namely, (1,0) + (1,0) + (1,0) + (1,0) → (10,0). It should be noted that the center hexagon of the base point is not unique. Alternatively, another center hexagon can be selected to build the base point of the Hex-ZZ BCS. As depicted in Figure S1b, we can still get the final chirality (10,0) via (6,0) + (1,-1)+ (1,-1) + (0,1) + (0,1) + (1,0) + (1,0) → (10,0). Obviously, the selection of the former center hexagon is much easier to determine the chirality with fewer shit steps because the arrangement of the six pentagons has a higher symmetry. In Figure S1c, by selecting the Pen-ZZ BCS, we can predict the (9,2) chirality via (5,0) + (1,0) + (0,1) + (0,1) + (1,0) +(1,0) + (1,0) → (9,2). In Figure S1d,e, we can get (8,3) chirality by selecting either Pen-AC BCS or Hex-ZZ BCS. Similarly, we can get (6,6) chirality by selecting three different BCS (see Figure S1f-h).

Based on these examples, we herein propose an algorithm that can "see" the chirality of the nascent carbon cap without knowing the helicity of the SWCNT wall: 1) Select a suitable BCS from Pen-AC, Pen-ZZ, Hex-AC, and Hex-ZZ for reference. Theoretically, the selection of the BCS, center pentagon or hexagon will not change the final chirality result. Nevertheless, just like the selection of the cartesian coordinate system, spherical coordinate system and cylindrical coordinate system in the mathematics, selecting the right BCS can make the algorithm simple and straightforward. The principle is that the six pentagons should be located on the extension lines of the selected BCS as much as possible; 2) Starting from the base point cap of the selected BCS, draw the shift route and shift steps of all the pentagons from their original positions to their current positions. According to the Figure 1k,l, label the chirality change for each shift step of the pentagon; 3) By summing all the chirality changes created by the shift steps and the base-point chirality, we can get the final chirality or the counterpart handedness. The advantages of this algorithm are: 1) we can directly "see" the chirality of the nascent cap at the early nucleation stage just according to the relative positions of the six pentagons without the elongation of the tube wall; 2) we can accurately know the corresponding chiral change for any shift of the pentagon during the cap nucleation; 3) we can distinguish the left- and right-handed chiralities; and 4) we can also transverse this algorithm to the computer and enumerate all the possible caps for a certain chirality.

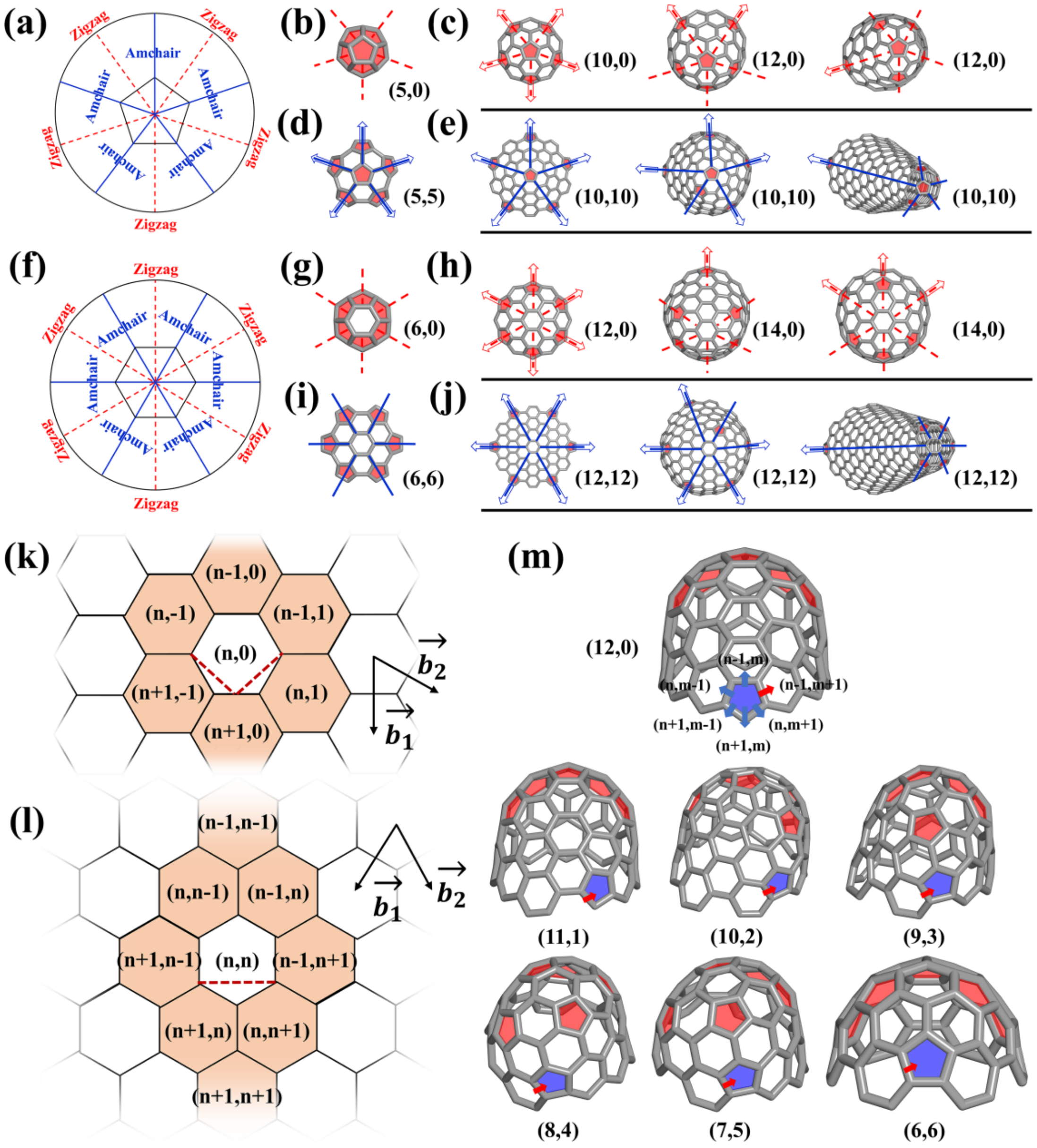

**Figure 1. Algorithm identifying cap chirality via pentagon shift**. (a) Pentagon-center BCS. (b,c) The base-point cap of the Pen-ZZ BCS (b) and the derived chiralities (c). (d,e) The base-point cap of (d) the Pen-AC BCS and (e) the derived chiralities. (f) Hexagon-center BCS. (g,h) The base-point cap of the (g) Hex-ZZ BCS and the (h) derived chiralities. (i,j) The base-point cap of (i) the Hex-

AC BCS and (j) the derived chirality. (k,l) Chirality evolution of one pentagon in (k) ZZ and (l) AC maps after moving to the surrounding hexagon position. (m) Gradual unidirectional evolution of the last pentagon (marked in violet) from (12,0) to (6,6) chirality cap.

### 3.2 Chirality mutation via the pentagon shift during the nascent cap nucleation

In the experiments, the near-AC SWCNTs, such as (n,n-1), (n,n-2) SWCNTs, are usually observed as the most abundant chirality during the selective synthesis. Many theories have been proposed to explain such an abundance.[62,67,68] All these theories focused on the dominated role of the SWCNT-catalyst interface but neglected the mutation of the carbon cap. Based on the algorithm mentioned above, the shift of the pentagon of the nascent cap during the nucleation stage may also contribute to the abundance of near-AC SWCNTs' production. To verify this assumption, we firstly take the (6,6) caps with $D_1$ and $C_{6v}$ symmetry as the prototypes for study. As shown in Figure 2, in order to promise the feasibility of the pentagon shift, the nascent cap with only one adder layer of carbon atom out of the pentagon are designed as the initial structure because the deeply embedded pentagons have little chance to move due to the steric hindrance. According to the chirality map in Figure 1l, three directions of the pentagon shift, namely, transverse shift, inward shift, and outward shift, are considered to mimic the chirality changes of (6,6) → (7,5) (Figure 2a), (6,6) → (6,5) (Figure 2b), and (6,6) → (7,6) (Figure 2e). As shown in Figure 2a,b, the transverse shift of pentagon can be finished simply by the 90 ° rotation of the C-C bond, which were predicated to override an activation barrier of ~ 2.27 eV on the metal catalyst.[81] Our calculation shows that there are a 0.3 eV (blue marker in Figure 2b) and 0.96 eV (blue marker in Figure S2b) energy rise for the $D_1$ (6,6) cap and $C_{6v}$ (6,6) cap on the $Ni_{55}$ particle after the transverse shift of the pentagon. The larger energy rise of the latter one may be mainly attributed to the adjacent pentagons formed that does not satisfy the isolated pentagon rule (IPR) (Figure S2a,b). The energy penalty of the adjacent pentagons on the nickel particle can be therefore estimated to be ~ 0.66 eV, which is smaller than ~1.5 eV calculated in the pure fullerene.[52] As shown in Figure 2c-f, considering the complicated atomic rearrangements for the inward shift and outward shift of the pentagons, including multiple carbon bond breakings and reconnections, we predesigned a possible kinetic route artificially with the relative energy profiles of the intermediates presented. And, the largest energy uphill in the energy profile is approximated as the energy barrier of the corresponding process. For the inward shift of the pentagon, there is an energy barrier of ~ 1.61 eV (red marker in Figure 2d) for the $D_1$ cap (Figure 2c,d) and ~ 2.26 eV (red marker in Figure S2d) for the $C_{6v}$ cap (Figure S2c,d), respectively. For the former one in Figure 2c,d, despite satisfying the IPR condition, the inward shift of pentagon still resulted in an increase in the final structure energy of approximately 0.77 eV (blue marker in Figure 2d) . In contrast, the latter one shows an energy decrease of ~ 0.53 eV (blue marker in Figure S2d) after the inward shift of the pentagon despite the adjacent pentagons formed, indicating the higher energy compensation on the cap-$Ni_{55}$ interface surface. Based on the continuum model, we can firstly estimate the increased curvature energy owning to the pentagon shift. As shown in Figure 2c,d, for the $D_1$ cap, the original conelike one undergoes a shrink of the radius after the chirality change (6,6) → (6,5) with the curvature energy increased by $\Delta E_c \sim \left(\frac{0.021}{r_2^2} - \frac{0.021}{r_1^2}\right) \times N_C^D$, where $r_1$ and $r_2$ are the radiuses of the (6,6) and (6,5) SWCNTs, respectively. And, $N_C^D$ is the number of carbon atoms of the locally deformed that is ~ 30 (see atoms marked in yellow in Figure S3a). The $\Delta E_c$ is hence ~ 0.73 eV, which is quite close to the calculated energy rise of ~ 0.77 eV in Figure 2d (blue marker). However, as shown in Figure S3b, the $C_{6v}$ cap undergoes a transformation from the hemispherical cap with a Gaussian curvature $K > 0$ to a conelike cap with Gaussian curvature $K = 0$. As shown in Figure S2d, the change of the curvature energy can be estimated by ~$\Delta E_c \sim \left(\frac{0.021}{r_2^2} - \frac{0.0593}{r_1^2}\right) \times N_C^D$, where $r_1$ and $r_2$ are the radiuses of the (6,6) and (6,5) SWCNT, respectively. And, $N_C^D$ is the number of carbon atoms of the locally deformed that is ~ 12 (see atoms marked in yellow in Figure S3b). The $\Delta E_c$ is hence ~ -2.48 eV, the magnitude of which is ~ 1.95 eV smaller than the calculated energy decrease of final structure in Figure S2d, i.e., $\Delta E$ ~ -0.53 eV (blue marker). Such a difference between the two energy decreases may be created by neglecting the energy penalties of the formation of the adjacent pentagons ($\Delta E$ ~ -0.66 eV) near the cap-$Ni_{55}$ interface calculated above and the formation of the dangling carbon bond ($\Delta E$ ~ 1.18 eV) created after the pentagon shift.[82] Figure 2e,f and Figure S2e,f show the outward shift of the pentagon on the $D_1$ cap and $C_{6v}$ (6,6) cap, respectively. The outward shift of the pentagon needs the assistance of the supplement of the other carbon atoms. As illustrated in Figure 2e,f, we considered a three-atoms chain and a mono adsorbed atom nearby to finish the outward shift. Firstly, the pentagon is transformed to the hexagon by the insertion of the carbon chain with an energy increase of only 0.47 eV (red marker in Figure 2e). Then, the carbon chains swing left and form a pentagon with dangling carbon atom attached. The final structure leads to an energy decrease of ~ - 0.11 eV (blue marker in Figure 2e). Similarly, for the $C_{6v}$ (6,6) cap (Figure S2e,f), the same process of the outward shift of the pentagon has an energy barrier of 0.79 eV(red marker in Figure S2e). The energy decrease of the final structure is ~ 0.03 eV (red marker in Figure S2e).

Herein, the pentagon shifts via the three routes either in the $D_1$ cap or the $C_{6v}$ cap are energetically feasible because of the moderate kinetic barriers and/or the thermodynamically preferred final transformed structures. And, the conelike $D_1$ cap requires lower energetic barriers than the hemisphere cap for the three routes of the pentagon shifts. Thus, the chirality mutation through the pentagon shift of the nucleated carbon cap is energetically feasible and can be expected to be a significant way to produce (n+1,n-1), (n,n-1) and (n+1,n) SWCNTs.

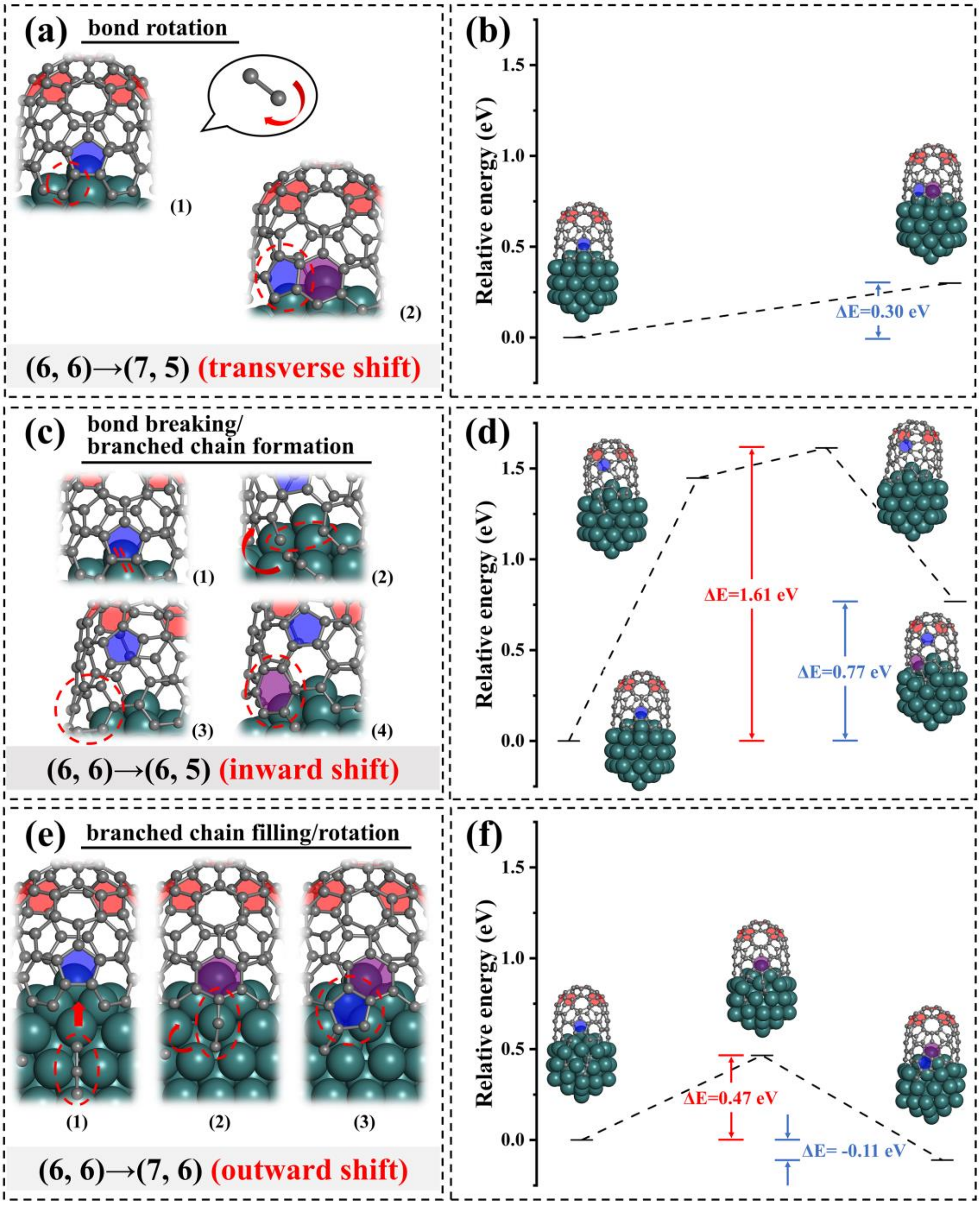

**Figure 2. Pentagon shift mechanism on the $D_1$ cap.** (a,b) Transition design route from (a) transverse shift and (b) the relative energy. (c,d) Transition design route from (c) inward shift and (d) the relative energy. (e,f) Transition design route from (e) outward shift and (f) the relative energy.

Besides, a significant discovery can be drawn is that the IPR does not necessarily work during the incipient cap evolution, either due to the much smaller $E_f$ on the cap-$Ni_{55}$ interface or due to the larger energy compensation induced by the adjacent pentagons, especially on the hemisphere cap nucleated on the small catalyst particles. In order to verify this discovery, we performed a series classic molecular dynamics (MD) simulation of SWCNT growth on the Ni particles, including $Ni_{19}$, $Ni_{32}$, and $Ni_{55}$. The MD simulations shows that there are ~ 53% non-IPR caps and ~ 47% IPR caps nucleated with varieties of identified chiralities (Figure 3), which confirms that the IPR does not fully determine the topological structures of these nucleated caps.

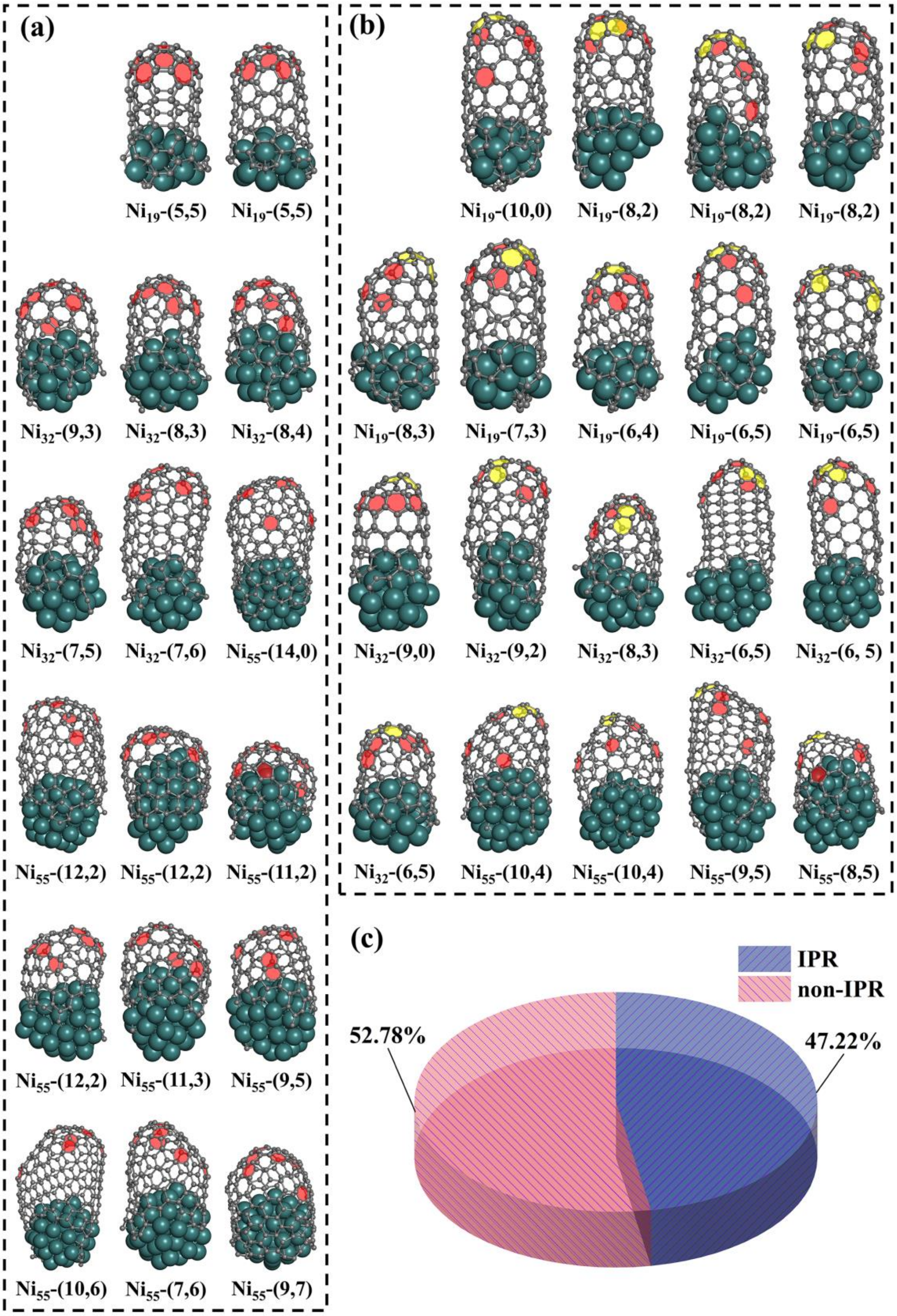

**Figure 3. Caps nucleated on the nickel particles based on the hybrid molecular dynamics simulation and basin hopping strategy.** (a) IPR caps. (b) non-IPR caps. (c) Statistical distribution chart of IPR and non-IPR.

### 3.3 Stability of short cap on the flat surface and curved particle

Then, we turned to investigate the dependence of the short cap stabilities on their topological geometries and the catalyst curvature. Without loss of generality, we select the caps of (12,0), (11,1), (9,3), (8,4), (7,5), and (6,6) for consideration, the (n,m) of which meets the n + m = 12 equation. Based on the algorithm above, we built four representative caps for each chirality that had different degree of symmetries. In order to determine the critical size of energetic stability, we calculated the energies of the (12,0) and (6,6) caps on the Ni(111) surface as a function of the number of carbon atoms (Figure S4). The critical sizes of the two caps reaching thermostability are roughly estimated at $N_c$ ~ 80. For this reason, each short cap is designed by adding one "layer" of hexagonal carbon surrounding the outmost pentagon. As shown in Figure 4a-f, we considered total 24 caps ranging from ZZ to AC on Ni(111) surface and $Ni_{55}$ particle, respectively. The $E_f$ and $E_{if}$ of the cap structures after optimization are illustrated in Figure 4g,h. The $E_f$ are within window of 0.18 eV/atom ~ 0.30 eV/atom, which is consistent with the average formation energies of (0.29 ± 0.01) eV/atom in the previous study.[52]

Overall, the carbon caps are mostly asymmetric structures with significant deformations upon contacting with the metal surfaces. Only a few chiralities, such as (12,0), (6,6), have the caps with high symmetry and their open ends are close to a symmetric shape. Thus, even for the same chirality, the cap-catalyst interfaces are quite different for the caps with different symmetries. It is clearly seen that the lowest $E_f$ of ZZ and AC carbon caps are lower compared with those of other chiralities both on Ni(111) and $Ni_{55}$ with the energetic difference ~ 0.02 eV/atom (or say ~ 1.2-1.8 eV). The lowest formation energy of AC carbon cap is also ~ 0.02 eV/atom lower than that of ZZ on Ni(111). On the $Ni_{55}$, however, their energy difference is almost negligible. For each chirality, the $E_f$ depends on the number of total carbon atoms, which decreases with the increase of the number of atoms ($E_f$ ~ $1/N_c$). Since the small number of the carbon atoms during the cap nucleation, the dispersion of the $E_f$ for each chirality is mostly larger than the ones of the elongated tubes, which will be discussed latter. As for the $E_{if}$ of the caps on the Ni(111), the (6,6) caps have the global minimum of the $E_{if}$ while the caps of the other chiralities have higher ones. The $E_f$ show a non-linear relation $E_f \sim \sin\theta$. In contrast, on the $Ni_{55}$, there is a clear negative linear correlation between the $E_{if}$ and $\theta$, i.e., $E_{if} \sim -\theta$. The dispersion of the $E_{if}$ for caps of each chirality on $Ni_{55}$ is much smaller than that on the Ni(111). Such a small dispersion can be attributed to the tangential contact between the cap edge and the $Ni_{55}$ particle. Thus, we can conclude here that the AC caps are still energetically favored among all of these caps and the Ni(111) surface can lower both the $E_f$ and $E_{if}$ via providing more atomic sites for interface matching because of the perpendicular contact between the cap and the Ni(111) surface.

When the number of carbon atoms is small, the energy of a carbon cap with more carbon atoms is always lower than that with a less carbon atoms, resulting in an inability to compare the structural stabilities between different carbon caps. Particularly, the effect of the cap symmetry cannot be "seen" in both the $E_f$ and $E_{if}$. To better reflect the different elastic deformations induced by the caps with different geometrical symmetries, we calculated the deformation energies ( $E_{defor}$ ) (see the $E_{defor}$ definition in Figure S5) of (12,0) and (6,6) carbon caps on Ni(111) and $Ni_{55}$ catalysts, which are defined as energetic difference of the carbon caps before and after contacting with the catalyst surface. As shown in Figure 4i, the $E_{defor}$ of (6,6) carbon caps are generally lower than that of (12,0) carbon caps after contacting with the metal catalysts. And, the $E_{defor}$ of the caps contacting with the $Ni_{55}$ are mostly smaller than those contacting with the Ni(111) except the (6,6)-I cap. The two AC caps, namely, (6,6)-I cap and (6,6)-II cap, have ultralow $E_{defor}$ on Ni(111) and $Ni_{55}$, respectively. A puzzling phenomenon is that the high symmetry of (12,0)-I cap fails to show advantages in reducing the $E_{defor}$ even on the Ni(111) surface that has a six-fold symmetry lattice. In addition, the (6,6)-I caps with $C_{6v}$ symmetry, transforms to an elliptical shape with $C_2$ symmetry that reduced the $E_{defor}$ efficiently. Such a puzzle should be understood by considering the interface match between the cap edge and the Ni(111) surface. Figure 4j-l shows the distributions of bond length, bond angle, and dihedral angle of the carbon-carbon (C-C) bonds at the (12,0) and (6,6) cap edges that contact the Ni(111) surface. Clearly, the C-C bonds of all the four (12,0) caps near the Ni(111) surface are mostly stretched to ~ 1.50 – 1.54 Å, largely deviating from the standard C-C bond length of 1.42 Å (Figure 4j). In contrast, the C-C bonds of the (6,6) caps near the Ni(111) surface are only stretched to ~ 1.44 – 1.47 Å, suggesting the smaller bond stretches. The bond angle distribution shows that the C-C bond angles are mostly compressed to be below the standard 120 ° except those of (6,6)-I structure (Figure 4k). For the (6,6)-I structure, the C-C bond angles distributed symmetrically around the 120 ° with the average angle ~ 121 °, indicating its small deformation penalty. Different from the bond stretch and bond angle distributions that reflecting the axial deformation, the dihedral angle distribution mainly reflects the radial deformation. As shown in Figure 4l, the (12,0)-I and (12,0)-III cap have narrow dihedral angle distributions, indicating their small radial deformations. The (12,0)-II and (12,0)-IV caps have broader dihedral angle distributions due to their obvious radial deformations. However, the larger radial deformations of (12,0)-II and (12,0)-IV have not resulted in their higher deformation energies. On the contrary, the axial deformation mainly contributes to the deformation penalty. Similarly, despite the broad dihedral angle distributions, all of (6,6) caps still have smaller $E_{defor}$ than (12,0) caps on the Ni(111) surface. Particularly, the (6,6)-I cap with ultralow $E_{defor}$ in Figure 4h prefers to sacrifice its high $C_{6v}$ symmetry by radial deformation to reduce its axial deformation when contacting the Ni(111) surface. This anomalous relationship can be explained by the radial flexibility of the SWCNT that will be discussed latter. As for the caps on the $Ni_{55}$ particle, in addition to the lattice matching of the interface, the size constrain of nanoparticle leads to less deformations of these caps. As shown in Figure S6, the bond lengths of the ZZ caps have been mostly reduced to ~1.44 Å. A small part of the bond lengths of the AC caps has been reduced to ~ 1.44 – 1.48 Å. The bond angles of the ZZ caps have been further compressed compared with the ones on the Ni(111) surface (Figure S7). However, the bond angles of all AC caps show a symmetric distribution around ~ 120 °, indicating the cancelled strain (Figure S8). The shorter bond lengths and symmetric bond angles therefore make the lower $E_{defor}$ for all AC caps on the $Ni_{55}$ particle.

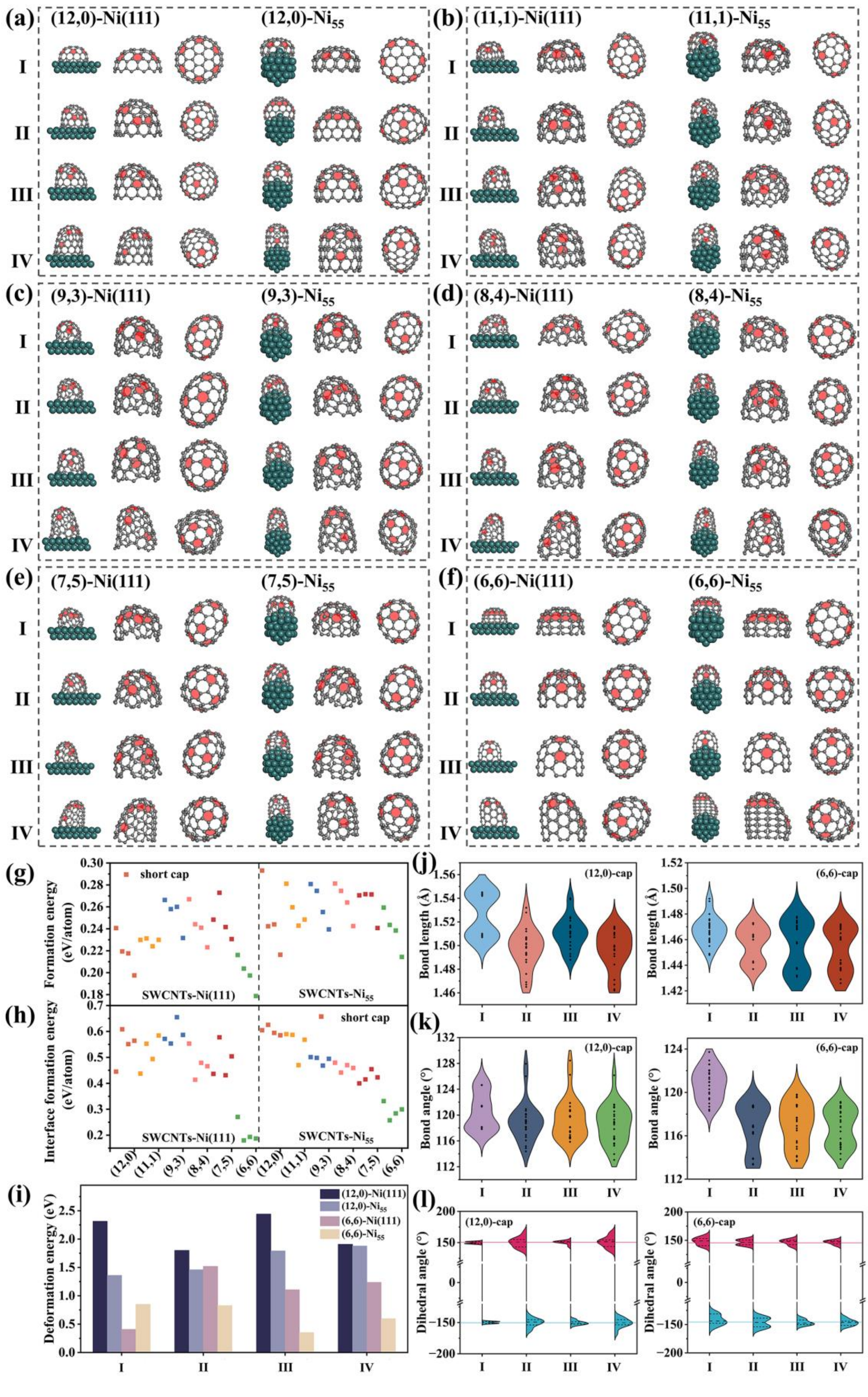

**Figure 4. Configurations and energies of different short caps for six chiral species on Ni(111)**

**and $Ni_{55}$ catalysts.** (a-f) Different views of short caps for six chiral species optimized on catalysts. (g,h) $E_f$ (g) and $E_{if}$ (h) distribution of short caps for six chiral species. (i) Deformation energies of (12,0) and (6,6) short caps. (j-l) The statistical distribution of (j) interfacial C-C bond lengths, (k) bond angles, and (l) dihedral angles of (12,0) and (6,6) short caps optimized on Ni(111).

### 3.4 Dual deformations of the cap and interface in the elongated SWCNTs

Next, we turned to consider the SWCNT elongated from the 24 short caps above on Ni(111) surface and $Ni_{55}$ particle with their optimized structures shown in Figure 5a-f. Inherited the symmetry of the short caps, the elongated SWCNTs present different deformation degrees of the interfacial edges for each chirality. In the previous theoretical studies, the SWCNT model without cap had been mostly used, with one open end or both ends usually passivated by the hydrogen atoms.[82-87] There should be only one most stable SWCNT-catalyst interface existing for each chirality.

However, once the capped SWCNT is considered, the edge of the open end for each chiral SWCNT is no longer a unique one owning to their different intrinsic deformations. There should be multiple intrinsic SWCNT-catalyst interfaces for the same chirality. Figure 5g-i compared the $E_f$, $E_{if}$, and $E_{in}$ of the short caps and elongated SWCNTs on both Ni(111) and $Ni_{55}$. As mentioned above, the $E_f$ of the SWCNTs are much lower than the ones of the caps on both Ni(111) and $Ni_{55}$ with the dispersion coefficient decreased greatly owing to their larger number of carbon atoms (Figure 5j). And, for the same chiral SWCNTs, the $E_f$ on Ni(111) are mostly lower than those of $Ni_{55}$. SWCNTs on Ni(111) show a greater energetic preference for the achiral species than the ones on $Ni_{55}$. Similar to the short caps, the elongated SWCNTs also show nonlinear and linear $E_{if}$ as a function of $\theta$ on Ni(111) and $Ni_{55}$, respectively, indicating that the independence of the $E_{if}$ on the SWCNT's length. The dispersion of the $E_{if}$ remains to be quite large on the Ni(111) but very small on the $Ni_{55}$ (Figure 5k). Based on the nonlinear relation $E_{if} \sim \sin(\theta)$ on the Ni(111) surface, we fitted the average $E_{if}$ as a function of $\theta$, namely,

$$E_{if} = |\gamma| \sin\left(\frac{\pi(\theta+c)}{w}\right) \qquad (5)$$

where $|\gamma|$ = 0.585 eV, $c$ = 12.6 °, and $w$ = 48 °. Similarly, the lowest $E_{if}$ as a function of $\theta$ on the $Ni_{55}$ can be fitted as

$$E_{if} = -\lambda\theta + \xi \qquad (6)$$

where $\lambda$ = 0.00948 eV/° and $\xi$ = 0.61111 eV.

Very interestingly, the $E_{in}$, indicating the $E_f$ of the bulk phase of the SWCNT, show an identical nonlinear increase as a function of $\theta$ both for caps and their elongated SWCNTs. Moreover, the relative relations of the short caps and elongated SWCNTs for each chirality are almost exact same on Ni(111) and $Ni_{55}$. These identical trends reflect the fact that the $E_{in}$ of the short cap or elongated SWCNT is largely determined by their curvatures instead of the interface. Herein, we propose a function which can fit the $E_{in}$ Vs. $\theta$ for either short caps or elongated SWCNTs (Figure 5i).

$$E_{in} = \frac{\varepsilon\pi^2\left(1-\beta tan^2\left(\frac{\pi}{180}\theta\right)\right)}{n^2a^2\left(1+\frac{4\tan\left(\frac{\pi}{180}\theta\right)^2}{(\sqrt{3}-\tan(\frac{\pi}{180}\theta))^2}+\frac{2\tan(\frac{\pi}{180}\theta)}{(\sqrt{3}-\tan\frac{\pi}{180}\theta)}\right)} \qquad (7)$$

Where, $\varepsilon = 14.50\, eV/Å^2$ , $\beta = 0.33$ , and $a = 2.46$ Å for short cap and $\varepsilon = 10.67\, eV/Å^2$ , $\beta = 0.36$ , and $a = 2.46$ Å for elongated SWCNT on Ni(111). Similarly, $\varepsilon = 16.60\, eV/Å^2$ , $\beta = 0.30$ , and $a = 2.46$ Å for short cap and $\varepsilon = 11.66\, eV/Å^2$, $\beta = 0.39$, and $a = 2.46$ Å for elongated SWCNT on $Ni_{55}$. Here, n is one of the chirality index (n,m) from 12 to 6. The detailed derivation of equation (7) can be found in the Supporting Information.

The dotted line calculated by pulsing the functions of equation (5), equation (6), and equation (7) can fit well the $E_f$ in Figure 5g very well. Obviously, the $E_f$ in Figure 5g is a balance result between the $E_{if}$ (Figure 5h) and the $E_{in}$ (Figure 5i). The details of the coefficients of the fitted curves are given in Table S1 in the Supporting Information. As we mentioned before, due to the different cap structures for each chirality, the elongated SWCNTs are not identical cylinders but the ones with different deformations. Ideally, the SWCNTs without caps for each chirality had only one $E_{in}$. However, there is a fluctuation within ~ 0.01 eV/atom for the $E_{in}$ of the different capped-SWCNTs, which comes from the different caps. Such a difference can be reflected by the similar relative trend between the $E_{in}$ and the $E_f$ for each chirality. For example, the relative trend of the $E_{in}$ of the capped SWCNTs for each chirality on $Ni_{55}$ has the almost same relative trend with their $E_f$ in Figure 5g. Besides, there is ~ 0.01 eV/atom energetic difference produced by the different cap structures. Despite the small energetic difference for each interior atom, the total energy difference is expected to be 0.01 eV/atom × 150 atom ~ 1.5 eV. That means, the total $E_{in}$ difference will be accumulated with the elongation of the SWCNTs. Such an accumulated energy bias will finally compete with the $E_{if}$ bias with the elongation of the caps. In another word, at the nucleation stage, when the carbon atom number is small, the $E_{if}$ dominates the nucleation probability. However, with the further elongation of the SWCNT, the $E_{in}$ will be a nonnegligible term that attend the energy bias together with the $E_{if}$ during the SWCNT growth. Among all these SWCNTs, the (6,6) SWCNT with cap-I structure has both lower $E_f$ and $E_{in}$ (Figure S9). However, the (12,0) SWCNT with cap-I structure is not. This means the capped (6,6) SWCNT with highest symmetry will dominate the $E_f$ compared with other SWCNTs with lower cap symmetries.

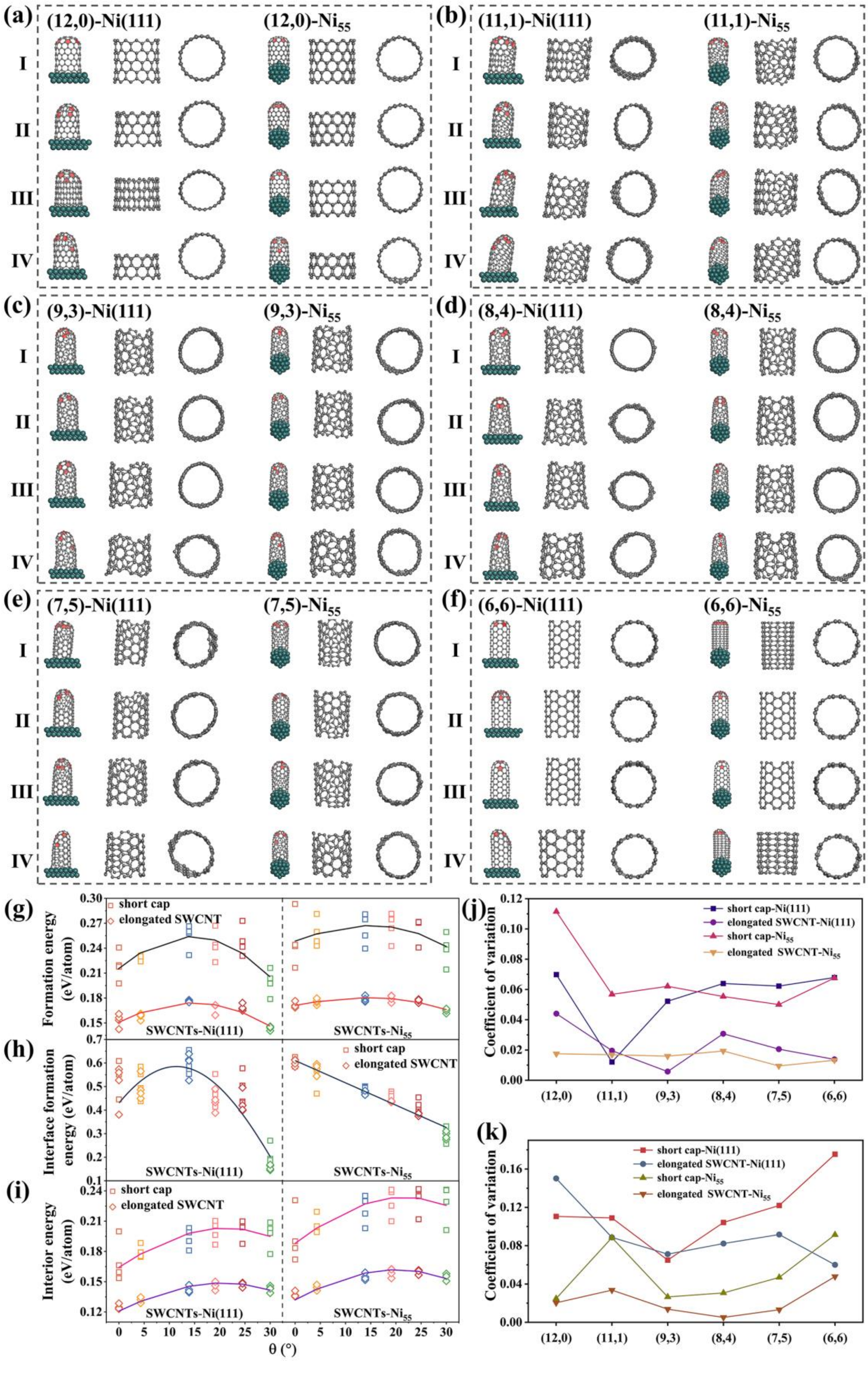

**Figure 5. Configurations and energies of different short caps and elongated SWCNTs on**

**$Ni(111)$ and $Ni_{55}$ catalysts.** (a-f) Different views of elongated SWCNTs optimized on Ni(111) and $Ni_{55}$ surface. (g) $E_f$. (h) $E_{if}$. (i) $E_{in}$. (j) Coefficient of variation of $E_f$. (k) Coefficient of variation of $E_{if}$.

As shown in Figure 6a, in order to characterize the deformation degree of the capped-SWCNTs, we take the upper and bottom carbon rings near the carbon cap and the metal catalyst as reference to define the shape factor ($S$) by the following equation (8):

$$S = \left(\frac{\frac{D_L^u}{D_S^u}+\frac{D_L^b}{D_S^b}}{2}\right) \cdot \left(\frac{d^u+d^b}{2}\right) \quad (8)$$

Where, the $D_L^u$ ($D_L^b$) and $D_S^u$ ($D_S^b$) are the long and short diameters of the deformed upper (bottom) carbon rings. The $d^u$ ($d^b$) is the projected distance between the two ends of $D_L^u$ ($D_L^b$) and $D_S^u$ ($D_S^b$). The magnitude of the $S$ therefore represents the deformation degree of each capped-SWCNT wall. As shown in Figure 6b,c, there is a quasi-liner correlation between the $E_f$ and the $S$ for each chirality SWCNT both on Ni(111) and $Ni_{55}$, i.e., $E_f \sim k * S$, where $k$ represents the slope of the fitted line segment, suggesting that the dual deformation of the capped-SWCNT originated from both the cap and the interface can indeed lead to an increase in the energy of the grown SWCNT. Figure 6d reveals the coefficient $k$ for each chirality on the Ni(111), the $k$ values of both (6,6) and (7,5) SWCNTs are almost half of the other chiralities. On the $Ni_{55}$, the $k$ of (11,1), (9,3), and (8,4) is much higher than the ones of the other chiralities. Among them, the $k$ values of (6,6) and (7,5) SWCNTs are always smaller, which implies that (6,6) and (7,5) SWCNTs are intrinsically flexible and thermodynamically advantageous in the catalytic growth process.

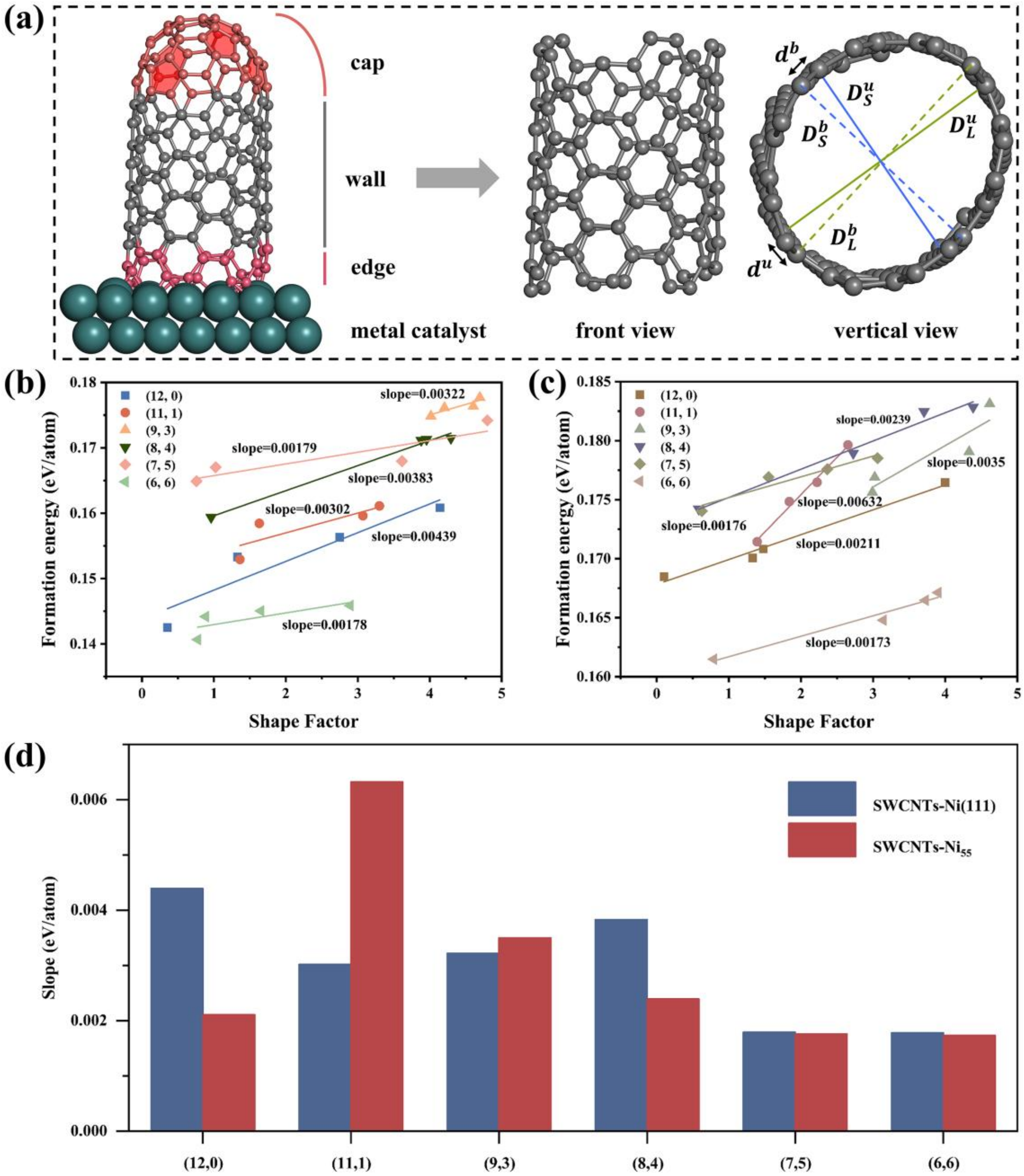

**Figure 6. Schematic diagram of SWCNT deformation and $S$.** (a) Schematic diagram of the wall deformation in a long SWCNTs. (b,c) Relationship between $E_f$ and $S$ of different chiral SWCNTs on (b) Ni(111) and (c) $Ni_{55}$. (d) The slope of the fitted line segment of $E_f$ Vs. $S$ on Ni(111) and $Ni_{55}$.

## 4. Conclusions

In this work, the influence of the cap structures on the SWCNTs nucleation and growth has been fully studied. During the stage of nascent cap nucleation, a novel algorithm based on the pentagons' shift has been proposed to describe the evolution of the cap structure and identify the chirality. The calculated energetic profiles of three pentagon-shift routes, namely transverse shift, inward shit, and outward shift, have confirmed the kinetic and thermodynamic feasibilities of the nascent cap evolutions on the metal catalyst with the chirality's changes of (n,n) → (n+1,n-1), (n,n)→(n,n-1), and (n,n)→(n+1,n), respectively. The IPR in the formation of the fullerenes was found to be not necessarily obeyed owing the larger energic compensation from the transformation of spherelike cap to conelike cap.

The energetic analysis in terms of $E_f$, $E_{if}$, and $E_{in}$ revealed that the topological structures of the different caps can indeed introduce energetic differences to the short caps and their derived SWCNTs owing to the dual-deformation effects coming from the cap and the carbon-catalyst interface, respectively. It was found that the ultra-high symmetry of topological structure does not guarantee its global minimum energy of the cap as well as its derived SWCNT. Only the AC cap with $C_{6v}$ symmetry shows a robust thermodynamic stability, either on the Ni(111) surface or on $Ni_{55}$ nanoparticle. For other chiralities, the Ni(111) and the $Ni_{55}$ can bias the relative energies of both $E_f$ and the $E_{if}$. The $E_{in}$ of the SWCNTs exhibits highly similar relative distributions on both Ni(111) and $Ni_{55}$, confirming that the energetic differences originates from the dual-deformation effect. The defined shape factor relates the dual-deformation degree to the $E_f$ of the SWCNT and reveals a perfect linear relation $E_f \sim k * S$. The AC SWCNT or near-AC SWCNTs were found to be highly flexible compared with other chirality SWCNTs under the dual-deformation effect. Thus, the robust thermostability of the AC cap, the feasible chirality mutation during the nascent cap nucleation ((n,n) →(n+1,n-1), (n,n)→(n,n-1), and (n,n)→(n+1,n)) via the pentagon shift as well as the higher flexibility of AC and near AC SWCNTs provide us a novel understanding on the high abundance of AC and near-AC SWCNTs synthesized in the catalytic growth.

**Acknowledgments**

This work was supported by the National Natural Science Foundation of China (12374180).

Supplementary Materials of

# Mysterious Role of Cap Configuration in Single-Walled Carbon Nanotube Catalytic Growth

Tianliang Feng, Ziwei Xu*

*School of Materials Science and Engineering, Jiangsu University, Zhenjiang 212013, China*

* Corresponding Author. E-mail address: ziweixu2014@ujs.edu.cn (Z. Xu)

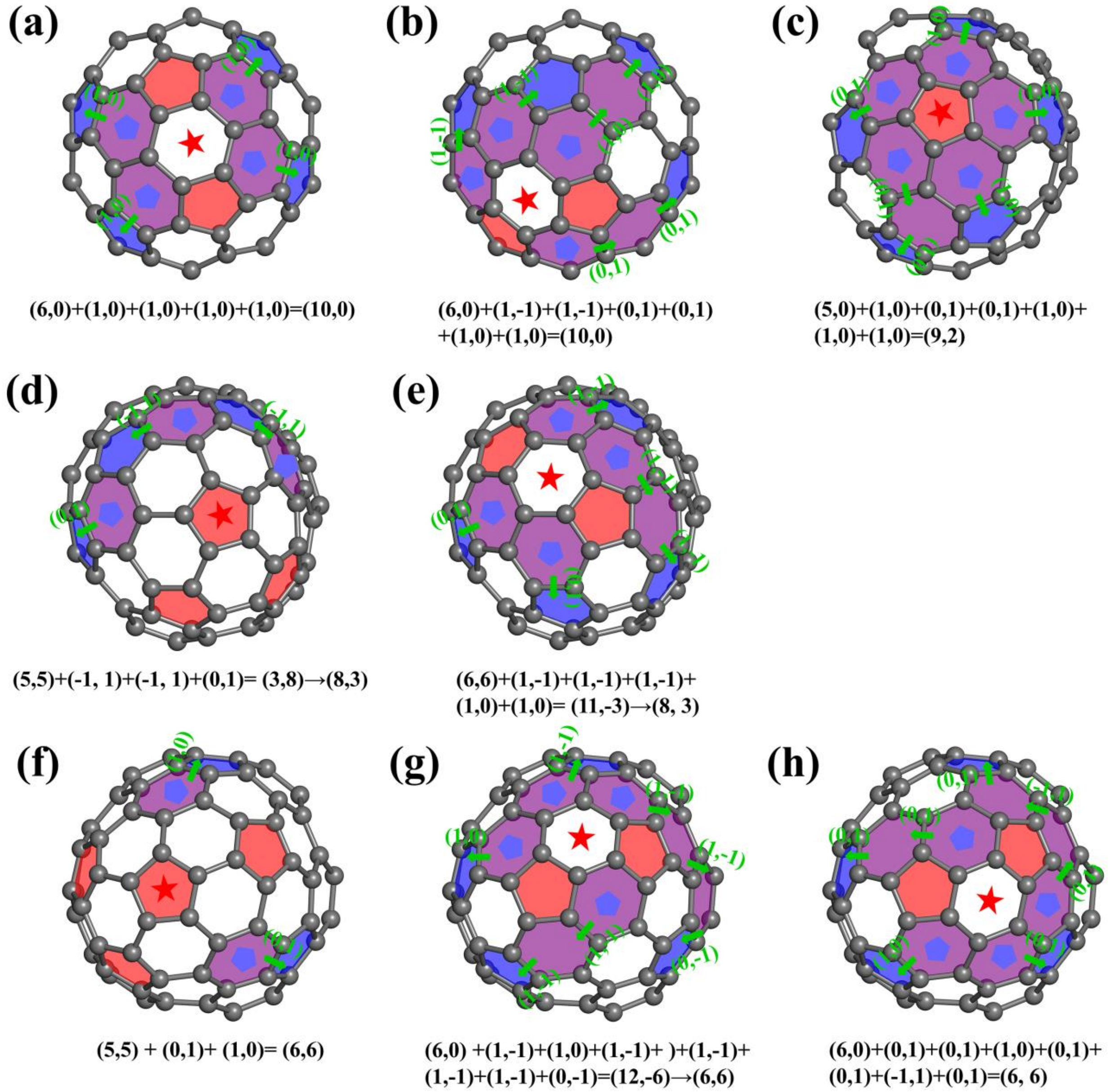

**Figure S1.** Examples of chirality evolution of nascent caps determined by basic coordiante system (BCS) and the pentagon shifts: (a,b) (10,0), (c) (9,2), (d,e) (8,3), and (f-h) (6,6). The red pentagram marks the center ring (pentagon or hexagon) surrounded by the pentagons of the selected BCS. The orange-filled rings represent the surrounding unmoving  pentagons and the small blue pentagons mark the origial positions of the surrounding shiftted pentagons. Each of the surrounding shiftted pentagons move to its final postion along the route guided by the green arrows with the passed hexagonal rings purple full-filled. The values on the green arrows represent change of the chirality index for each shift step.

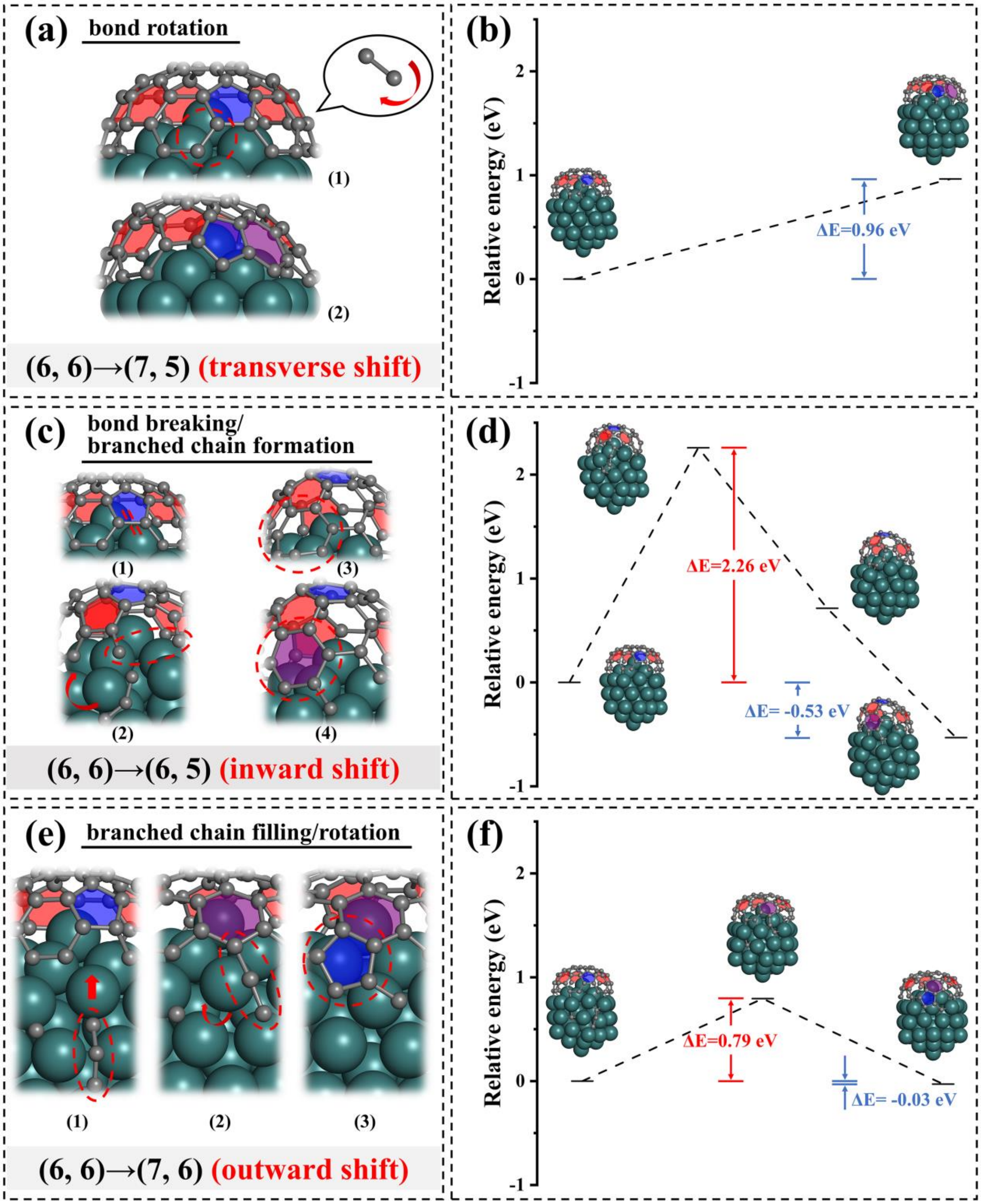

**Figure S2. Pentagon shift mechanism on the $C_{6v}$ cap.** (a,b) (n,n) Transition design route (a) from transverse shift and (b) the relative energy; (c,d) Transition design route from (c) inward shift and (d) the relative energy; (e,f) (n,n) Transition design route (e) from outward shift and (f) the relative energy.

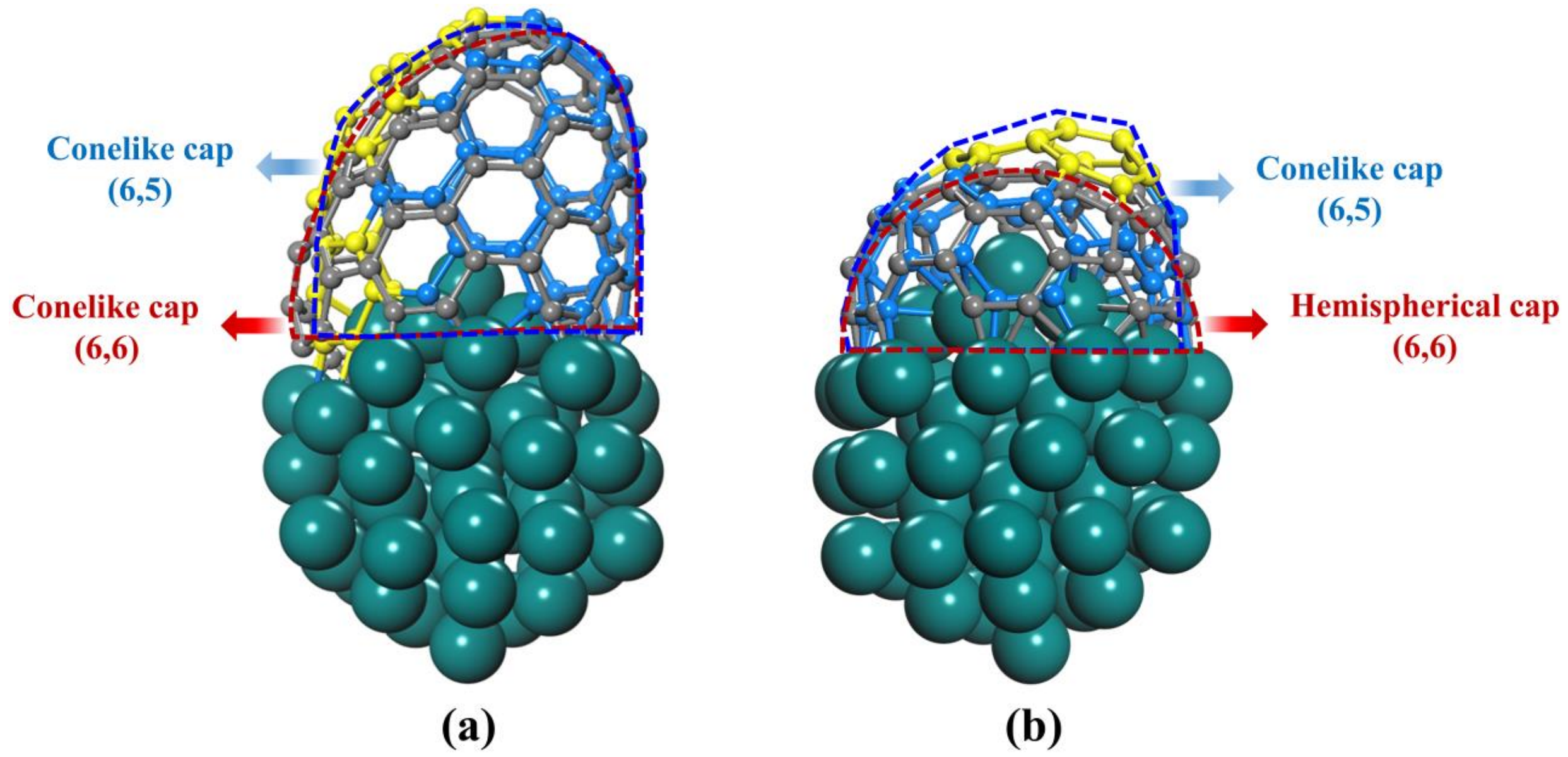

**Figure S3. Deformed caps before and after the inward shift of the pentagon.** (a) conelike (6,6) cap → conelike (6,5) cap, (b) hemispherical (6,6) cap → conelike (6,5) cap. The dark green balls represent the nickel atoms. The gray and blue balls represent the carbon atoms of caps before and after the pentagon shift, respectively. Atoms exhibiting pronounced deformation following pentagon shift are highlighted in yellow, with atomic numbers approximately (a) 30 and (b) 12.

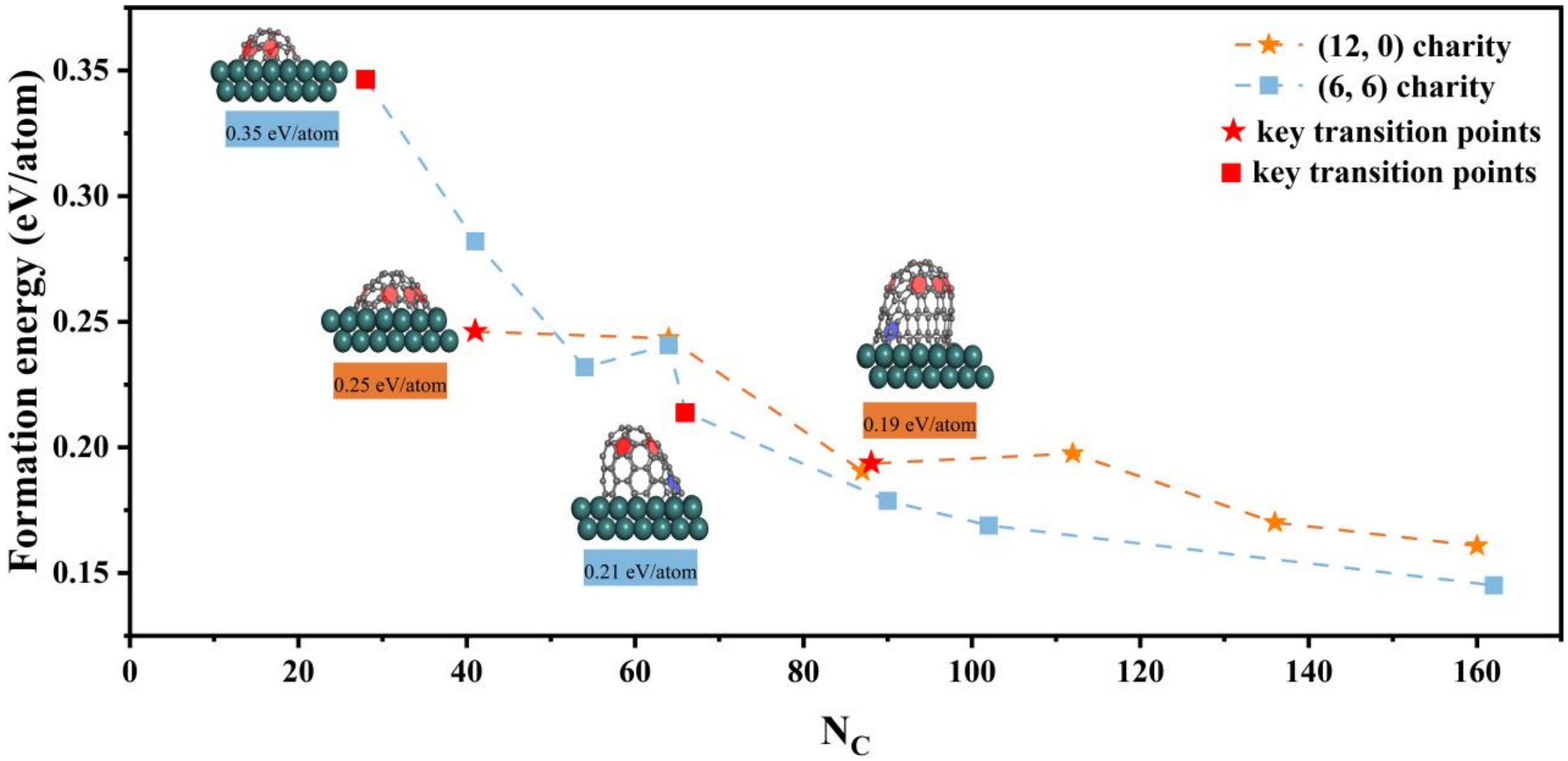

**Figure S4. The formation energy of the (12,0) and (6,6) caps on the Ni(111) surface as a function of the number of carbon atoms.**

The method used to calculate the deformation energy ($E_{defor}$) is illustrated in Figure S5. The (12,0) carbon cap on a Ni(111) catalyst is used here as an example. The single-point energy of the (12,0) carbon cap optimized for hydrogen passivation on the Ni(111) catalyst is denoted as $E_1$ in Figure S5a. The single-point energy of the (12,0) carbon cap optimized to form a full fullerene after hydrogen passivation, by using half of the fullerene structure is denoted as $E_2$ in Figure S5b. The $E_{defor}$ is calculated by the following equation (S1):

$$E_{defor} = E_1 - E_2 \qquad \text{(S1)}$$

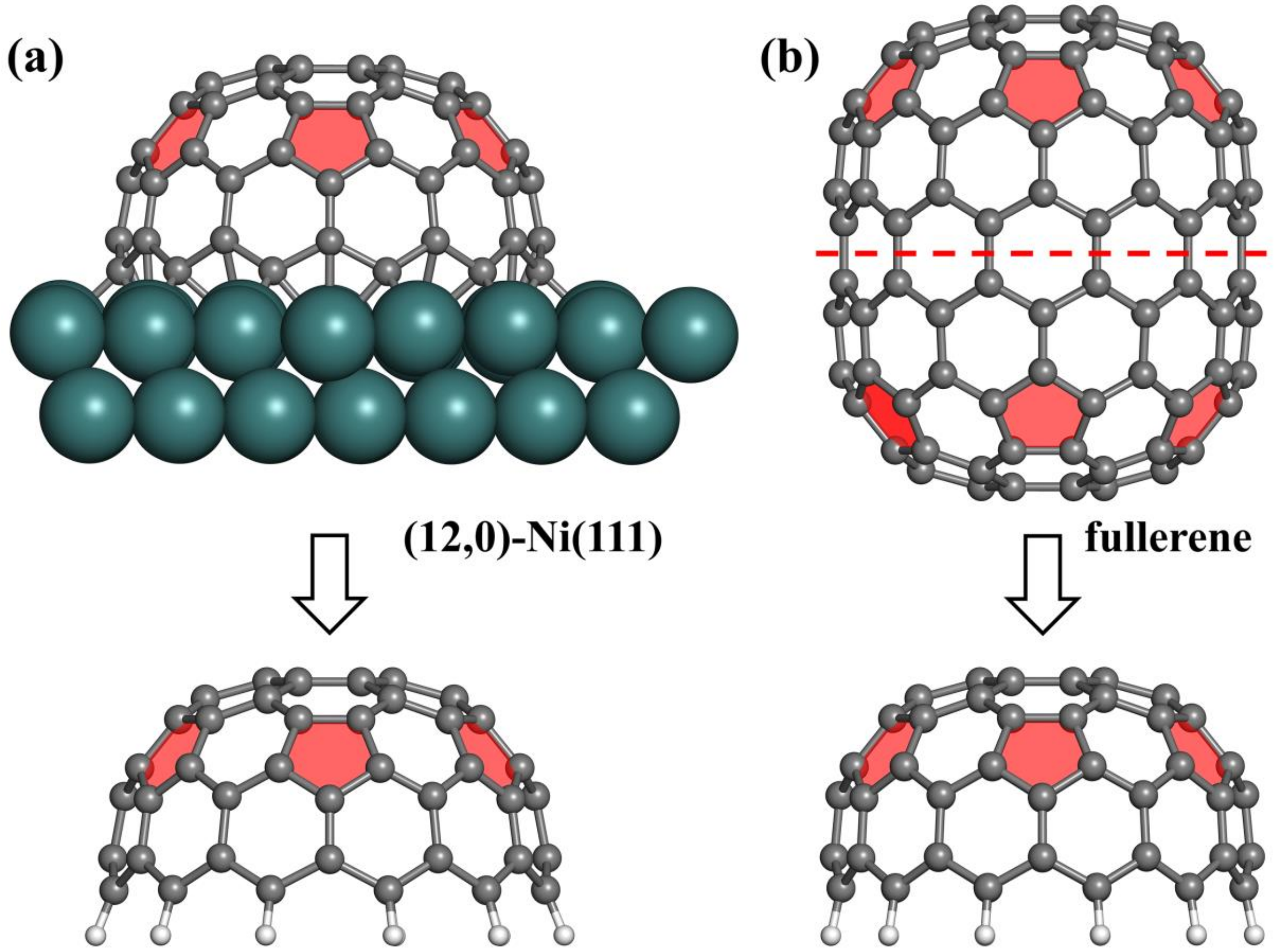

**Figure S5. Schematic definition of deformation energy.** (a) (12,0) carbon caps placed over Ni(111) catalyst optimized for hydrogen passivation. (b) (12,0) carbon caps were placed over a Ni(111) catalyst optimized to synthesize a complete fullerene, and then half of the fullerene was taken for hydrogen passivation.

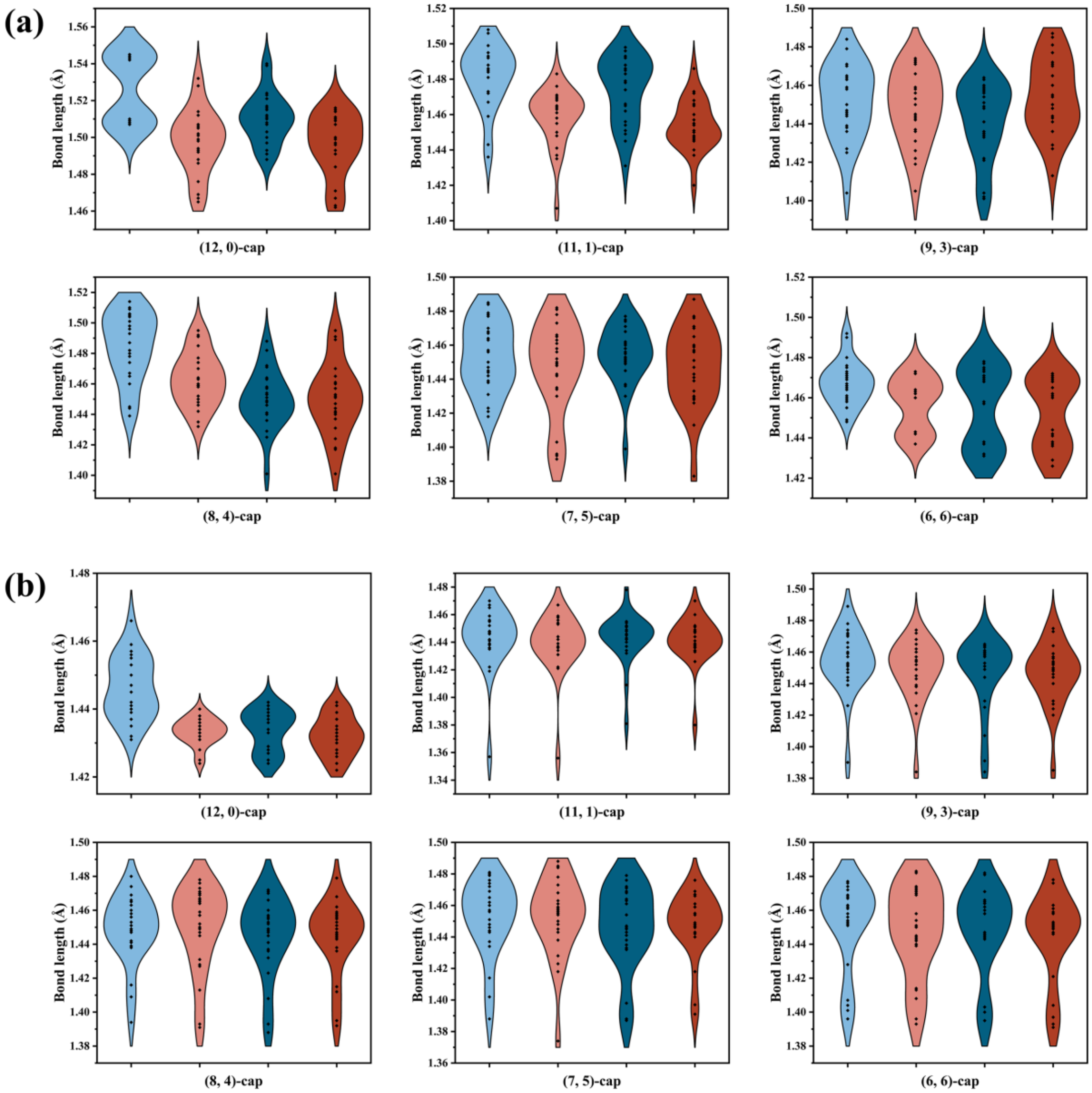

**Figure S6.** C-C bond length distributions at the interface of different chiral short cap structures optimized on (a) Ni(111) plane and (b) $Ni_{55}$ particle.

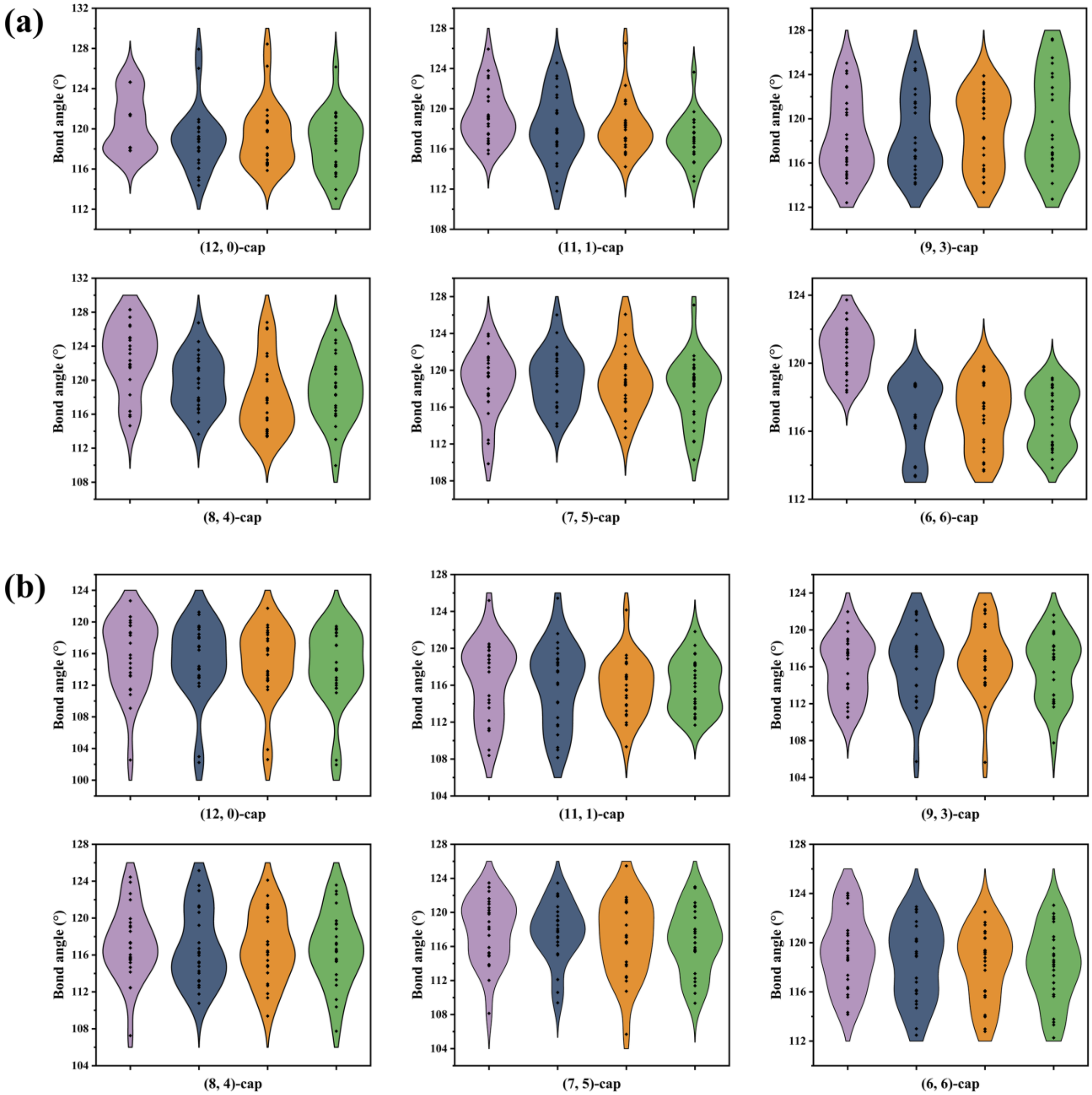

**Figure S7.** C-C bond angle distributions at the interface of different chiral short cap structures optimized on (a) Ni(111) plane and (b) $Ni_{55}$ particle.

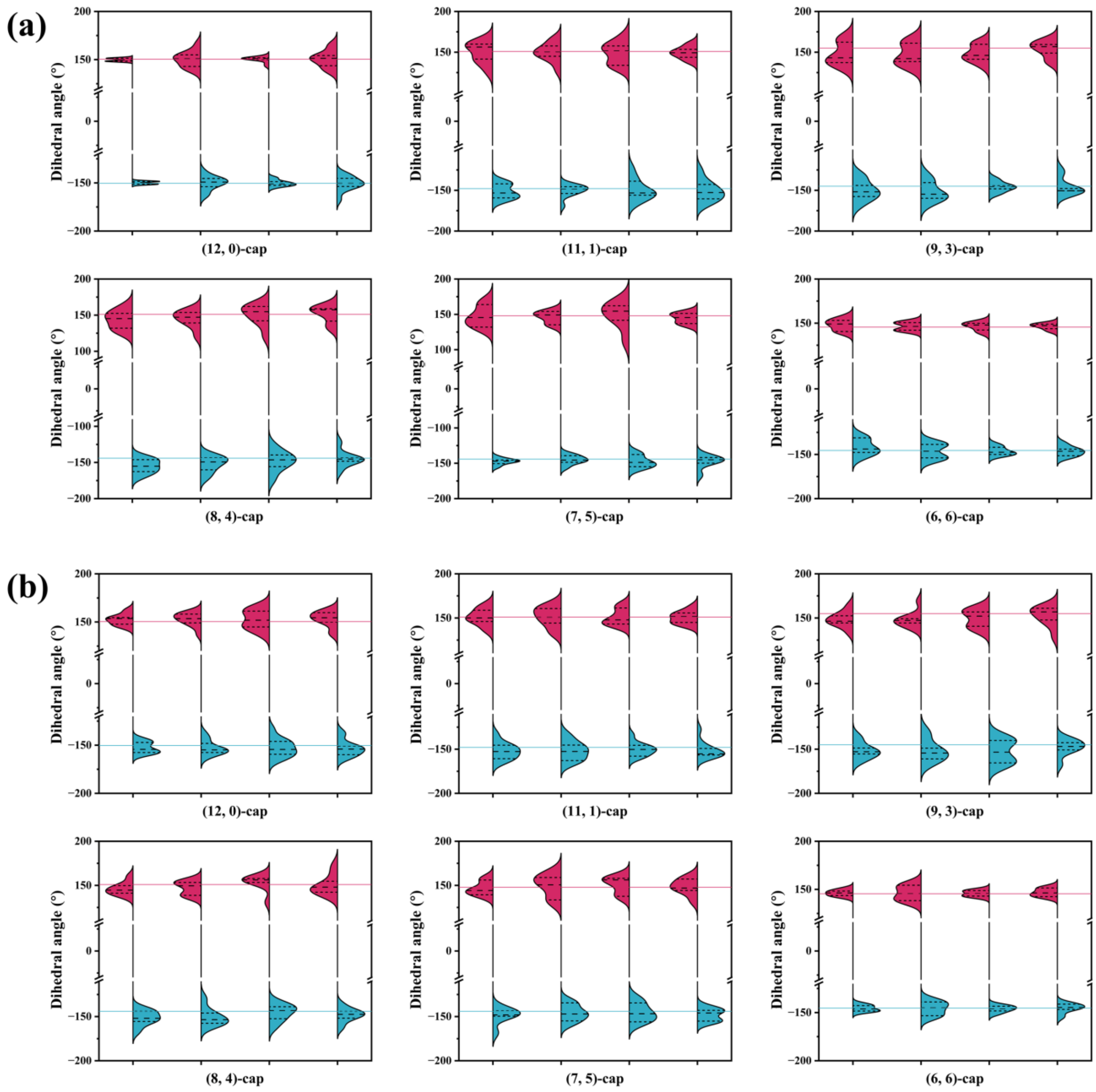

**Figure S8.** C-C dihedral angle distributions at the interface of different chiral short cap structures optimized on (a) Ni(111) plane and (b) $Ni_{55}$ particle.

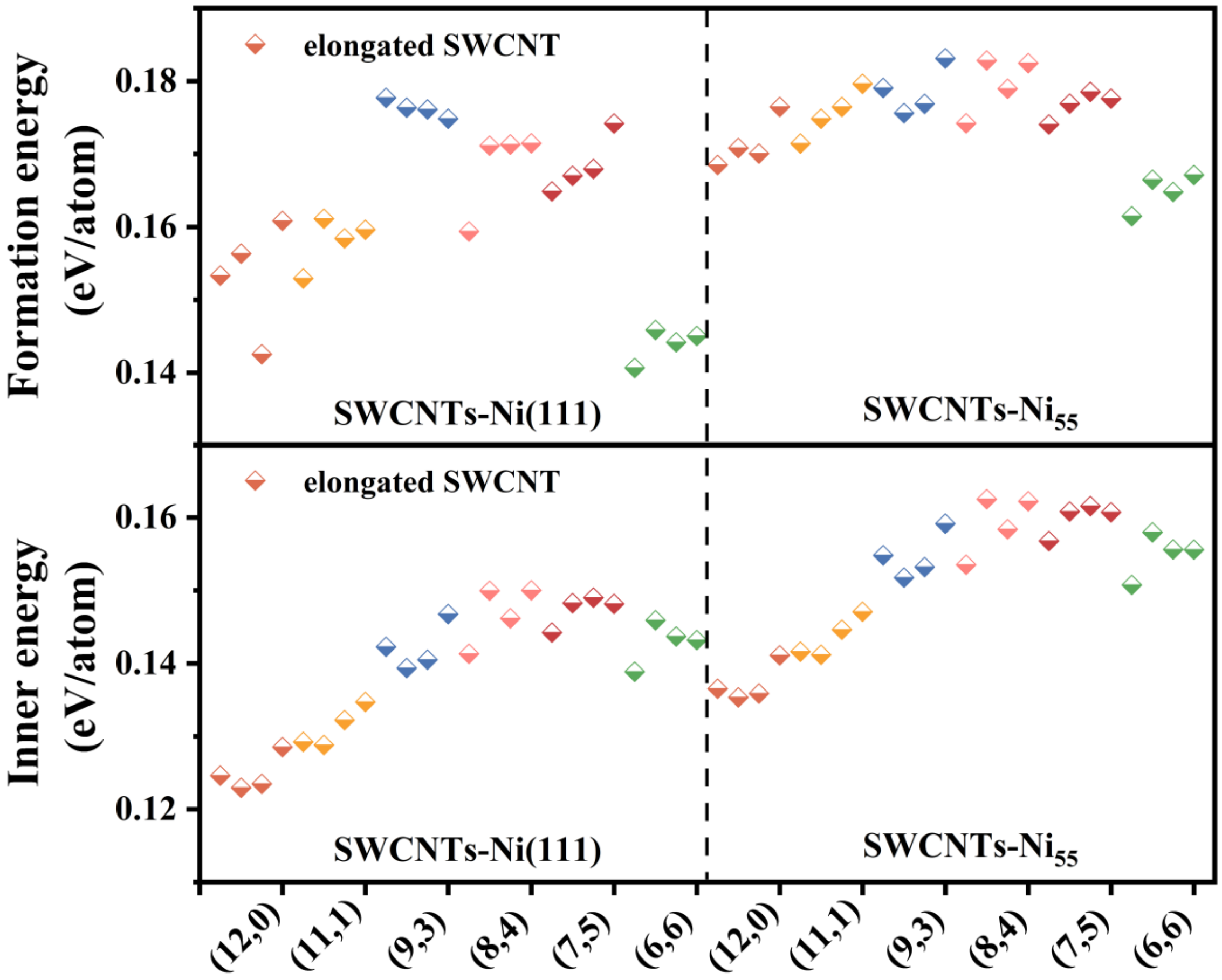

**Figure S9. Formation energy and inner energy distribution of elongated SWCNTs on Ni(111) and $Ni_{55}$.**

**Table S1.** The coefficients of the fitted curves in Figure 4(g-i).

| | formation energy | | | | interface formation energy | | inner energy | | | |
|---|---|---|---|---|---|---|---|---|---|---|
| | Ni(111) | | $Ni_{55}$ | | Ni(111) | $Ni_{55}$ | Ni(111) | | $Ni_{55}$ | |
| | short cap | elongated SWCNT | short cap | elongated SWCNT | Ni(111) | $Ni_{55}$ | short cap | elongated SWCNT | short cap | elongated SWCNT |
| $\lvert\gamma\rvert$/ eV | 0.1 | 0.1 | / | / | 0.59 | / | / | / | / | / |
| $c$/ ° | 21.89 | 28.22 | / | / | 12.6 | / | / | / | / | / |
| $w$/ ° | 67.77 | 82.49 | / | / | 48 | / | / | / | / | / |
| $\lambda$/ eV/° | / | / | 1.72*10^-11 | 2.07*10^-8 | / | 0.0094 | / | / | / | / |
| $\xi$/ eV | / | / | 0.14 | 0.11 | / | 0.61 | / | / | / | / |
| $\varepsilon$/ eV/Å ^2 | 11.49 | 5.52 | 10.0014 | 5.15 | / | / | 14.5 | 10.67 | 16.6 | 11.66 |
| $\beta$ | 0.6 | 0.6 | 0.88 | 0.94 | / | / | 0.33 | 0.36 | 0.3 | 0.39 |
| $a$/ Å | 2.46 | 2.46 | 2.46 | 2.46 | / | / | 2.46 | 2.46 | 2.46 | 2.46 |

**Detailed derivation of equation (7) in the main text:**

According to the relation between SWCNT diameter and the chiral index (n,m), the diameter of the SWCNT can be calculated by the following equation (S2):

$$D = a\sqrt{n^2+m^2+nm}/\pi \qquad \text{(S2)}$$

Where a is the lattice parameter of the carbon wall ($a = 0.246$ nm).

According to the definition of chiral angle, we have

$$\tan(\frac{\pi}{180}\theta) = \frac{\sqrt{3}m}{2n+m} \qquad \text{(S3)}$$

According to equation S2 and equation S3, we hence have

$$D = \frac{na}{\pi}\sqrt{1+\frac{4tan^2\left(\frac{\pi}{180}\theta\right)}{(\sqrt{3}-\tan(\frac{\pi}{180}\theta))^2}+\frac{2\tan(\frac{\pi}{180}\theta)}{(\sqrt{3}-\tan(\frac{\pi}{180}\theta))}} \qquad \text{(S4)}$$

Based on the curvature energy $E_{cur} \sim \varepsilon/D^2$,we have the interior energy $E_{in}$

$$E_{in}\sim\frac{\varepsilon\pi^2}{n^2a^2(1+\frac{4\tan\left(\frac{\pi}{180}\theta\right)^2}{(\sqrt{3}-\tan\frac{\pi}{180}\theta)^2}+\frac{2\tan(\frac{\pi}{180}\theta)}{(\sqrt{3}-\tan\frac{\pi}{180}\theta)})} \qquad \text{(S5)}$$

Where the $\varepsilon$ is the parameter of energy per atom fitted based on the energy profile.

In order to fit the energy of $E_{in}$ vs. $\theta$ in Figure 5i better, we add a correction term,

$$E_{in} = \frac{\varepsilon\pi^2(1-\beta tan^2\left(\frac{\pi}{180}\theta\right))}{n^2a^2(1+\frac{4\tan\left(\frac{\pi}{180}\theta\right)^2}{(\sqrt{3}-\tan(\frac{\pi}{180}\theta))^2}+\frac{2\tan(\frac{\pi}{180}\theta)}{(\sqrt{3}-\tan\frac{\pi}{180}\theta)})} \qquad \text{(S6)}$$

Where, $(1-\beta tan^2\left(\frac{\pi}{180}\theta\right))$ is the correction term with the parameter $\beta$ fitted by $E_{in}$ vs. $\theta$.